\documentclass[prd,twocolumn,nofootinbib,preprintnumbers,superscriptaddress]{revtex4}

\usepackage[colorlinks,linkcolor={blue},citecolor={blue},urlcolor={red}]           {hyperref}

\usepackage{graphicx}

\usepackage{amsmath}

\usepackage{cleveref}

\usepackage{upgreek}

\usepackage{xspace}

\usepackage[caption=false]{subfig}

\usepackage{multirow}

\usepackage{enumerate}

\usepackage[acronym,shortcuts,hyperfirst=true]{glossaries}
\glossarystyle{altlist}

\makeglossaries

\newacronym{AAAS}{AAAS}{American Association for the Advancement of Science}
\newacronym{acprobe}{AC}{acoustic probe}
\newacronym{AC}{AC}{alternating current}
\newacronym{ACC}{ACC}{accelerometer}
\newacronym{ACIGA}{ACIGA}{Australian Consortium for Interferometric Gravitational Astronomy}
\newacronym{ACWP}{ACWP}{actual cost of work performed}
\newacronym{ADC}{ADC}{analogue-to-digital converter}
\newacronym{ADE}{ADE}{Advanced Detector Era}
\newacronym{ADCU}{ADCU}{analogue data collection unit}
\newacronym{AdV}{AdV}{Advanced Virgo}
\newacronym{AEI}{AEI}{Albert Einstein Institute}
\newacronym{aLIGO}{aLIGO}{Advanced \protect\ac{LIGO}}
\newacronym{AM}{AM}{amplitude modulation}
\newacronym{AMU}{AMU}{atomic mass unit}
\newacronym{ANU}{ANU}{Australian National University}
\newacronym{AOC}{AOC}{adaptive optics compensation}
\newacronym{AOM}{AOM}{acousto-optic modulator}
\newacronym{AOS}{AOS}{auxiliary optics system}
\newacronym{API}{}{application programming interface}
\newacronym{AR}{AR}{anti-reflective}
\newacronym{AS}{AS}{antisymmetric}
\newacronym{AS1}{AS1}{antisymmetric port, PD 1}
\newacronym{AS2}{AS2}{antisymmetric port, PD 2}
\newacronym{AS3}{AS3}{antisymmetric port, PD 3}
\newacronym{AS4}{AS4}{antisymmetric port, PD 4}
\newacronym{ASC}{ASC}{alignment sensing and control}
\newacronym{ASD}{ASD}{amplitude spectral density}
\newacronym[see={AWGg}]{AWG}{AWG}{Arbitrary Waveform Generator}
\newacronym{BAT}{BAT}{Burst Alert Telescope}
\newacronym{BATSE}{BATSE}{Burst And Transient Source Experiment}
\newacronym{BBH}{BBH}{binary black hole}
\newacronym{BNS}{BNS}{binary neutron star}
\newacronym{BS}{BS}{beam splitter}
\newacronym{CARM}{CARM}{common arm}
\newacronym{CBC}{CBC}{compact binary coalescence}
\newacronym{CMB}{CMB}{cosmic microwave background}
\newacronym{CW}{CW}{continuous wave}
\newacronym{cwb}{cWB}{Coherent WaveBurst}
\newacronym{DARM}{DARM}{differential arm}
\newacronym{DetChar}{DetChar}{detector characterisation}
\newacronym[longplural={degrees of freedom}]{DOF}{DOF}{degree of freedom}
\newacronym{DMT}{DMT}{Data Monitoring Tool}
\newacronym{DQ}{DQ}{data quality}
\newacronym{E10}{E10}{\protect\ac{LIGO} Engineering Run 10}
\newacronym{E11}{E11}{\protect\ac{LIGO} Engineering Run 11}
\newacronym{E12}{E12}{\protect\ac{LIGO} Engineering Run 12}
\newacronym{E13}{E13}{\protect\ac{LIGO} Engineering Run 13}
\newacronym{E14}{E14}{\protect\ac{LIGO} Engineering Run 14}
\newacronym{E1}{E1}{\protect\ac{LIGO} Engineering Run 1}
\newacronym{E2}{E2}{\protect\ac{LIGO} Engineering Run 2}
\newacronym{E3}{E3}{\protect\ac{LIGO} Engineering Run 3}
\newacronym{E4}{E4}{\protect\ac{LIGO} Engineering Run 4}
\newacronym{E5}{E5}{\protect\ac{LIGO} Engineering Run 5}
\newacronym{E6}{E6}{\protect\ac{LIGO} Engineering Run 6}
\newacronym{E7}{E7}{\protect\ac{LIGO} Engineering Run 7}
\newacronym{E8}{E8}{\protect\ac{LIGO} Engineering Run 8}
\newacronym{E9}{E9}{\protect\ac{LIGO} Engineering Run 9}
\newacronym{EDR}{EDR}{efficiency-to-deadtime ratio}
\newacronym{EFE}{EFE}{Einstein Field Equation}
\newacronym{EM}{EM}{electromagnetic}
\newacronym{ER1}{ER1}{\protect\ac{ER} 1}
\newacronym{ER2}{ER2}{\protect\ac{ER} 2}
\newacronym{ER3}{ER3}{\protect\ac{ER} 3}
\newacronym{ER}{ER}{\protect\ac{aLIGO} Engineering Run}
\newacronym{etg}{ETG}{event trigger generator}
\newacronym{ETM}{ETM}{end test mass}
\newacronym{eLIGO}{eLIGO}{Enhanced \protect\ac{LIGO}}
\newacronym{FAP}{FAP}{false-alarm probability}
\newacronym{FAR}{FAR}{false-alarm rate}
\newacronym{FFT}{FFT}{fast Fourier transform}
\newacronym[longplural={figures of merit}]{FOM}{FOM}{figure of merit}
\newacronym{GBM}{GBM}{Gamma-ray Burst Monitor}
\newacronym{GCN}{GCN}{Gamma-ray Coordinates Network}
\newacronym{GEO-HF}{GEO-HF}{GEO High Frequency}
\newacronym{GEODC}{GEODC}{\protect\gls{GEO} detector characterisation}
\newacronym{GPS}{GPS}{Global Positioning System}
\newacronym{GRB}{GRB}{gamma-ray burst}
\newacronym{GR}{GR}{general relativity}
\newacronym{GWB}{GWB}{\protect\glsuseri{GW} burst}
\newacronym[user1={gravitational-wave}]{GW}{GW}{gravitational wave}
\newacronym{HACR}{HACR}{Hierarchical Algorithm for Curves and Ridges}
\newacronym{HEN}{HEN}{high-energy neutrino}
\newacronym{HEPI}{HEPI}{hydraulic external pre-isolation}
\newacronym{hveto}{HVeto}{HierarchichalVeto}
\newacronym{IFAR}{IFAR}{inverse \protect\ac{FAR}}
\newacronym{IMC}{IMC}{input \protect\acl{MC}}
\newacronym{IMR}{IMR}{inspiral-merger-ringdown}
\newacronym{IPN}{IPN}{InterPlanetary Network}
\newacronym{ISCO}{ISCO}{innermost stable circular orbit}
\newacronym{ITM}{ITM}{input test mass}
\newacronym{iLIGO}{iLIGO}{Initial \protect\ac{LIGO}}
\newacronym{KAGRA}{KAGRA}{the Kamioka Gravitational Wave Detector}
\newacronym{KW}{KW}{Kleine-Welle}
\newacronym{LAL}{LAL}{the \protect\ac{LIGO} Algorithm Library}
\newacronym{LCGT}{LCGT}{the Large-scale Cryogenic Gravitational Wave Telescope} 
\newacronym{LHO}{LHO}{\protect\ac{LIGO} Hanford Observatory}
\newacronym{LIGO}{LIGO}{Laser Interferometer Gravitational-Wave Observatory}
\newacronym{LLO}{LLO}{\protect\ac{LIGO} Livingston Observatory}
\newacronym{LSC}{LSC}{\protect\ac{LIGO} Scientific Collaboration}
\newacronym{LVEA}{LVEA}{Large Vacuum Equipment Area}
\newacronym{MC}{MC}{mode cleaner}
\newacronym{MCMC}{MCMC}{Markov-chain Monte Carlo}
\newacronym{MICH}{MICH}{Michelson degree of freedom}
\newacronym{NSNS}{NSNS}{neutron star-neutron-star}
\newacronym{NSBH}{NSBH}{neutron star-black hole}
\newacronym{ODC}{ODC}{Online Detector Characterisation}
\newacronym{OMC}{OMC}{output \protect\acl{MC}}
\newacronym{PDF}{PDF}{probability density function}
\newacronym{PEM}{PEM}{physical environment monitor}
\newacronym{PEPI}{PEPI}{piezoelectric external pre-isolation}
\newacronym{PN}{PN}{post-Newtonian}
\newacronym{PRC}{PRC}{\protect\acl{PR} cavity}
\newacronym{PRM}{PRM}{\protect\acl{PR} mirror}
\newacronym{PR}{PR}{power recycling}
\newacronym{PSD}{PSD}{power spectral density}
\newacronym{PyLAL}{PyLAL}{the Python \protect\ac{LAL}}
\newacronym{QPD}{QPD}{quadrant photo-diode}
\newacronym{rms}{rms}{root-mean-square}
\newacronym{RF}{RF}{radio frequency}
\newacronym{S1}{S1}{Science Run 1}
\newacronym{S2}{S2}{Science Run 2}
\newacronym{S3}{S3}{Science Run 3}
\newacronym{S4}{S4}{Science Run 4}
\newacronym{S5}{S5}{Science Run 5}
\newacronym{S6}{S6}{Science Run 6}
\newacronym{SEI}{SEI}{seismic isolation system}
\newacronym{SFT}{SFT}{short Fourier transform}
\newacronym{SGR}{SGR}{soft gamma repeater}
\newacronym{SGWB}{SGWB}{stochastic gravitational-wave background}
\newacronym{SNR}{SNR}{signal-to-noise ratio}
\newacronym{SRC}{SRC}{\protect\ac{SR} cavity}
\newacronym{SRM}{SRM}{\protect\ac{SR} mirror}
\newacronym{SR}{SR}{signal recycling}
\newacronym{TEM}{TEM}{transverse electro-magnetic}
\newacronym{TCS}{TCS}{thermal compensation system}
\newacronym{TT}{TT}{transverse, traceless}
\newacronym{UPV}{UPV}{Used Percentage Veto}
\newacronym{USA}{USA}{United States of America}
\newacronym{USDOE}{USDOE}{United States Department of Energy}
\newacronym{VSR}{VSR}{Virgo Science Run}
\newacronym{VDQ}{VDQ}{Virgo Data Quality}
\newacronym{VSR1}{VSR1}{Virgo Science Run 1}
\newacronym{VSR2}{VSR2}{Virgo Science Run 2}
\newacronym{VSR3}{VSR3}{Virgo Science Run 3}
\newacronym{VSR4}{VSR4}{Virgo Science Run 4}

\newglossaryentry{action}{
    name={action},    description={the mathematical law describing the motion of the system}}
\newglossaryentry{astrowatch}{
    name={astrowatch},    description={a program to archive science-quality data when                  commissioning activities do not disable the detector}}
\newglossaryentry{ATel}{
    name={ATel},    description={The Astronomer's telegram: a service for reporting new                  astronomical observations}}
\newglossaryentry{AWGg}{
    name={AWG},    description={system used to inject signals into the interferometer,                  including \protect\glspl{hwinj}},    see={AWG}}
\newglossaryentry{CAT1}{
    name={category 1},    description={\hl{FIXME}}}
\newglossaryentry{CAT2}{
    name={category 2},    description={\hl{FIXME}}}
\newglossaryentry{CAT3}{
    name={category 3},    description={\hl{FIXME}}}
\newglossaryentry{CAT4}{
    name={category 4},    description={\hl{FIXME}}}
\newglossaryentry{control loop}{
    name={control loop},    description={an electromechanical chain to measure the behaviour of a                  system and feedback a mitigating force to the input}}
\newglossaryentry{DC readout}{
    name={DC readout},    description={a special case of homodyne detection}}
\newglossaryentry{deadtime}{
    name={deadtime},    description={the fractional amount of analysis time that has been vetoed}}
\newglossaryentry{efficiency}{
    name={efficiency},    description={the fractional number of GW triggers removed by a veto}}
\newglossaryentry{Fermi}{
    name={Fermi},    description={NASA spacecraft in low-earth orbit, to study                  \protect\acp{GRB}}}
\newglossaryentry{GEO}{
    name={GEO600},    description={the British-German detector at Ruthe}}
\newglossaryentry{glitch}{
    name={glitch},    description={transient noise event},    plural={glitches}}
\newglossaryentry{harmonic gauge}{
    name={harmonic gauge},    description={a coordinate system whose coordinates $x_i$ all satisfy                  Laplace's equation}}
\newglossaryentry{hwinj}{
    name={hardware injection},    description={a signal inserted into the instrument by means of                  mechanically osciallating one of the end test masses in                  order to test understanding of the detector and the                  effectiveness and efficiency of detection algorithms}}
\newglossaryentry{light dip}{
    name={light dip},    description={drop in the power stored in a detector arm cavity},    descriptionplural={drops in the power stored in a detector arm cavity}}
\newglossaryentry{line}{
    name={spectral line},    description={long-duration, narrow-bandwidth noise peak},    descriptionplural={long-duration, narrow-bandwidth noise peaks}}
\newglossaryentry{lock}{
    name={lock},    description={the process of maintaining resonance in all necessary                  optical cavities required for sensitive detector operation},    user1={locked}}
\newglossaryentry{ltt}{
    name={light travel time},    description={time taken for a light beam to travel directly between                  two detectors}}
\newglossaryentry{multipole expansion}{
    name={multipole expansion},    description={An expansion in $(v/c)$ used to find approximate                  solutions to the linearised Einstein Field Equations},    see={post-Newtonian expansion}}
\newglossaryentry{null stream}{
    name={null stream},    description={a coherent combination of detector data that contains                  no potential \protect\ac{GW} power}}
\newglossaryentry{off-source}{
    name={off-source},    description={the time window used in a targeted search to describe the                  background noise}}
\newglossaryentry{on-source}{
    name={on-source},    description={the time window used in a targeted search to detect a                  \protect\ac{GW} signal from a \protect\ac{GRB}}}
\newglossaryentry{periastron}{
    name={periastron},    description={the point in a binary orbit at which the objects are                  closest together}}
\newglossaryentry{post-Newtonian expansion}{
    name={post-Newtonian expansion},    description={An expansion in $(v/c)$ used to find approximate                  solutions to the linearised Einstein Field Equations}}
\newglossaryentry{playground}{
    name={playground},    description={data from the full analysis time used to tune                  signal-consistency tests and signal-based                  \protect\glspl{veto}}}
\newglossaryentry{SeisVeto}{
    name={SeisVeto},    description={the \protect\ac{LIGO} seismic veto developed using                  targeted veto methods in \protect\ac{S6}}}
\newglossaryentry{sbv}{    name={signal-based veto},    description={removal from the analysis of an event trigger based on                  it's template-matching parameters, rather than instrumental                  vetoes},    plural={signal-based vetoes}}
\newglossaryentry{short GRB}{
    name={short \protect\acs{GRB}},    description={a \protect\ac{GRB} that emit the central 90\% of their                  energy in under 2\,seconds}}
\newglossaryentry{short-wave expansion}{
    name={short-wave expansion},    description={an expansion of the \protect\aclp{EFE} to separate                  short-wavelength \protect\aclp{GW} from a long-wavelength                  background}}
\newglossaryentry{software injection}{
    name={software injection},    description={a signal inserted into the data stream to test the                  effectiveness and efficiency of a detection algorithm}}
\newglossaryentry{Swift}{
    name={Swift},    description={NASA spacecraft in low-earth orbit, to study                  \protect\acp{GRB}}}
\newglossaryentry{targeted}{
    name={targeted},    description={a search for gravitational waves from an event detected                  by another method, such as a \protect\ac{GRB}}}
\newglossaryentry{template bank}{
    name={template bank},    description={a set of probable gravitational-wave signals used in the                  matched-filter algorithm}}
\newglossaryentry{trigger}{
    name={trigger},    description={event produced by a \protect\ac{GW} search algorithm},    descriptionplural={events produced by a \protect\ac{GW} search algorithm}}
\newglossaryentry{upconversion}{
    name={upconversion},    description={non-linear coupling of low-frequency noise into higher                  frequency/broadband noise}}
\newglossaryentry{weak-field approximation}{
    name={weak-field approximation},    description={a description of the curved-space metric as a perturbed                  state of the flat-space metric.}}
\newglossaryentry{veto}{
    name={veto},    description={time segment indicating poor \protect\gls{DQ} to be                  removed from an analysis},    descriptionplural={time segments indicating poor \protect\gls{DQ}                        to be removed from an analysis},    plural={vetoes}}

\glsdisablehyper

\AtBeginDocument{    \defglsdisplayfirst{\textit{#1#4}}
}

\newcommand{\varmathbf}[1]{\boldsymbol{\mathbf{#1}}}
\newcommand{\msun}{\ensuremath{\textrm{M}_\odot}\xspace}
\newcommand{\chisq}{\ensuremath{\chi^2}\xspace}

\newcommand{\sub}[1]{\ensuremath{_{\textrm{#1}}}}

\DeclareGraphicsExtensions{    .pdf,.PDF,    .png,.PNG,    .jpg,.mps,.jpeg,.jbig2,.jb2,.JPG,.JPEG,.JBIG2,.JB2}

\newcommand{\dccversion}{LIGO-P1500106}

\begin{document}

\title{A fully-coherent all-sky search for gravitational-waves from 
       compact binary coalescences}
\author{D. Macleod}
\affiliation{School of Physics and Astronomy, Cardiff University, Cardiff, UK}
\affiliation{Louisiana State University, Baton Rouge, LA 70803, USA }
\author{I. W. Harry}
\affiliation{School of Physics and Astronomy, Cardiff University, Cardiff, UK}
\affiliation{Department of Physics, Syracuse University, Syracuse NY}
\affiliation{Max Planck Institut f\"{u}r Gravitationsphysik, Am M\"{u}hlenberg 1, D-14476, Potsdam-Golm, Germany} 
\author{S. Fairhurst}
\affiliation{School of Physics and Astronomy, Cardiff University, Cardiff, UK}

\date{\today}
\preprint{\dccversion}
\begin{abstract}
We introduce a fully-coherent method for searching for gravitational wave signals generated
by the merger of black hole and/or neutron star binaries.  This extends the coherent analysis
previously developed and used for targeted gravitational wave searches to an all-sky, all-time
search.  We apply the search to one month of data taken during the fifth science run of the LIGO
detectors.  We demonstrate an increase in sensitivity of 25\% over the coincidence search,
which is commensurate with expectations.  Finally, we discuss prospects for implementing and
running a coherent search for gravitational wave signals from binary coalescence in the advanced
gravitational wave detector data.
\end{abstract}
 \maketitle

\glsresetall

\section{Introduction}
\label{sec:intro}

In recent years, the \ac{LIGO}~\cite{Abbott:2007kv} and Virgo \cite{Acernese:2007zze} have operated 
as a network of ground-based \ac{GW} detectors in an attempt to detect and study signals of 
astrophysical origin.  The data have been 
searched for evidence of gravitational waves from \acp{CBC}~\cite{Colaboration:2011np, 
Briggs:2012ce,Aasi:2013km},  unmodelled \ac{GW} bursts~\cite{Briggs:2012ce,Abadie:2012rq}, 
non-axisymmetric spinning neutron stars~\cite{Aasi:2012fw}, and a \ac{SGWB}~\cite{Abbott:2009ws}.
While these searches yielded no direct detections, great strides were made in both instrumental 
science and data analysis techniques, paving the way for highly-anticipated second-generation, or 
advanced, detectors~\cite{TheLIGOScientific:2014jea, Acernese:2015gua}.
The advanced LIGO detectors are expected to begin operation in late 2015, with Virgo joining a year 
or two later, and evolve to their design sensitivity over the following years \cite{Aasi:2013wya}. 
In addition, the KAGRA detector \cite{PhysRevD.88.043007} in Japan and a third LIGO detector in 
India \cite{Iyer:2011wb} are expected to join the global network.

In the coming years, the first direct observations of gravitational waves are expected 
\cite{Abadie:2010cf} and binary mergers of neutron stars and/or black holes are the most promising 
astrophysical sources.  It has long been argued that the most sensitive way to search for gravitational 
waves from a network of detectors is to use a coherent search 
\cite{Bose:1999pj, Finn:2000hj, Pai:2000zt} in which data from all detectors are combined
in an optimal way prior to performing the search.  While this method has been applied to searches for 
unmodelled burst sources  \cite{Abbott:2009zi},
it has proven more difficult to use for the binary merger search.  Instead, a coincidence search has 
been used \cite{Colaboration:2011np, Babak:2012zx}, whereby the data
from individual detectors are searched independently and the events recorded in the different 
detectors checked for time and mass
consistency appropriate for a gravitational wave signal.  

The benefit of the coincidence search is that it reduces the computational cost, at the 
expense of some loss in sensitivity.  The dominant cost of the search is in performing 
a matched filter of the data against a bank of template waveforms \cite{Allen:2005fk}.  
The coincidence search performs this task once per detector per template.  A search over the 
sky location of the signal is then trivially done by time-shifting the results from the different 
detectors accordingly \cite{Babak:2012zx}.
Similarly, the noise background is estimated by applying larger time shifts to the data 
(significantly longer than the light travel time between detectors) to search for noise coincidences.
A naive implementation of the coherent search would require an independent filtering of the data for 
each template and each sky point, with the search repeated for each time-shift used to estimate 
the noise background.  Computationally, this is not feasible.  

There are, however, good reasons to believe that the coherent search will be more sensitive
than the coincident one \cite{Finn:2000hj}, providing the motivation to overcome the 
computational challenges of the coherent analysis \cite{Pai:2000zt, Canton:2014ena}.  In a 
coincidence search, it is necessary to place a threshold on the \ac{SNR} of events which will
be stored by the analysis prior to identifying coincidences.  This means that the power in the
\ac{GW} signal will only be accumulated in those detectors where there was an event above
threshold.  In comparison, the coherent analysis naturally incorporates the \ac{SNR} from all
operational detectors.   In the coincidence analysis, each detector is searched independently,
and there is no guarantee that the observed signals in each detector are compatible with a 
\ac{GW} source with two polarizations.  In the coherent analysis the data from all detectors
is combined to extract the two physical \ac{GW} polarizations.  When there are more than two
detectors, it is then possible to generate a null stream (or streams)
\cite{GuerselTinto1989, Wen:2005ui} which will contain only noise.  Removing these additional
noise contributions from the \ac{SNR} will enhance the sensitivity of the search.  The 
benefit of a coherent analysis becomes more significant as the number of detectors in the
global network increases.  Thus, with the realistic prospect of a five detector network operating
in the next few years, there is increased motivation to overcome the challenges posed by a
coherent \ac{CBC} search.

In the past few years, a coherent search for gravitational waves associated with \acp{GRB} has
been developed \cite{Harry:2010fr, Williamson:2014wma} and used in analyses of
LIGO and Virgo data \cite{Briggs:2012ce, Aasi:2014iia}.  The observed \ac{GRB} signal is 
used to restrict the sky location and arrival time of the \ac{GW} signal, which significantly
reduces the computational cost of the analysis.  Nonetheless, in developing a coherent, targeted
search for binary merger signals, many of the issues involved in performing an all-sky, all-time
analysis have been addressed.  In particular, the analysis has been constructed so that 
each template is filtered independently through the data from each detector and the single detector
(complex) SNR time series are used to search over the sky and also perform a 
small number of time shifts.  Thus, the calculation of the single detector \ac{SNR} remains the 
dominant computational cost.  Furthermore, the algorithm was shown to improve search sensitivity 
by around 30\% when compared to a coincidence-based search over the same 
data~\cite{Harry:2010fr}.  In this work, we extend the targeted, coherent search to an all-sky search, 
demonstrating the first fully-coherent all-sky search for \acp{GW} from \acp{CBC}.  

This paper is laid out as follows.  In section \ref{sec:coherent}, we begin by briefly reviewing the coherent search 
as implemented in \cite{Harry:2010fr}, then we describe our new methods to
extend that work to an all-sky coherent search.  This requires
searching over amplitude parameters, sky position, time, component masses and spins.  In addition,
we briefly recap the signal consistency tests that are used to mitigate the effects of non-stationary
data.  In \ref{sec:two_site_search}, we describe in detail the search as performed with the 
three-detector, two-site LIGO network as it existed in LIGO's fifth science run, giving the results of the search in \ref{sec:all_sky_efficiency}.  We also discuss the prospects for a coherent analysis of
advanced detector data, and evaluate the likely sensitivity improvements.  We end, in section 
\ref{sec:discussion}, with a summary and discussion of future prospects. \section{A coherent search for coalescing binary systems}
\label{sec:coherent}

The gravitational waveforms emitted by a coalescing binary can be calculated by the 
post-Newtonian formalism when the two compact objects are well separated
\cite{Blanchet:2013haa}.  As
they spiral closer, higher order terms in the post-Newtonian expansion become increasingly
important and numerical relativity is used to calculate the waveform in the final stages
of inspiral and through the merger and ringdown of the merged system 
\cite{Hannam:2009rd, Centrella:2010}.  Using this
information, a number of phenomenological models have been constructed that accurately
describe the gravitational waveform over a large region of the parameter space of binaries
which do not precess (i.e. the components are either non-spinning or have spins aligned with
the orbital angular momentum) \cite{Santamaria:2010, Taracchini:2013}.  Consequently, when performing a search for these systems,
matched filtering techniques are generally used.  

The binary coalescence waveform for binaries in circular orbit
depends upon at least fifteen parameters, and possibly more
if we include the equation of state for neutron stars.  These parameters are the two masses, six
components of the spin (encoding magnitude and orientation of the two spins), the location 
(distance, right ascension and declination) and orientation (inclination, polarization and phase) 
of the binary, and the time of coalescence.  The search we describe below is restricted to binaries
where the spin-induced precession of the orbit can be neglected, for which the component 
spins must be aligned
with the binary's orbital angular momentum \cite{Santamaria:2010, Taracchini:2013}.  
This restricts the number of parameters to eleven.
In addition, we focus only on the dominant harmonic of the waveform and ignore higher modes \cite{VanDenBroeck:2006qu, Capano:2013raa, Varma:2014jxa}.  

In the rest of this section, we describe the method by which each of these 11 dimensions of the parameter
space is covered in the search.  Finally, we discuss additional features which have been developed to mitigate
the presence of non-gaussian artefacts in the detector data.

\subsection{Maximizing over amplitude parameters: $D, \iota, \psi, \phi$}
\label{sec:coherent_snr}

We begin by considering four parameters --- the distance, binary inclination, polarization and coalescence 
phase --- that affect only the observed amplitude and phase of the waveform in the various detectors.  
For the purposes of a coherent search, the amplitude of a \ac{GW} signal from a non-precessing binary 
inspiral can be decomposed into two polarisations~\cite{Harry:2010fr},
\begin{subequations}
\begin{align}
    h_+(t) &= \mathcal{A}^1h_0(t) + \mathcal{A}^3h_{\pi/2}(t), \\
    h_\times(t) &= \mathcal{A}^2h_0(t) + \mathcal{A}^4h_{\pi/2}(t) \, .
\end{align}
\end{subequations}
Here, $h_{0}(t)$ and $h_{\pi/2}(t)$ describe the two phases of the waveform, which depend upon 
the masses and spins of the binary components and are usually
assumed to be orthogonal.  The amplitudes $\mathcal{A}^{i}$ are
\begin{subequations}
\begin{align}
    \mathcal{A}^1 &= \frac{D_0}{D} \frac{\left(1 + \cos^2\iota\right)}{2}                          \cos2 \phi_0 \cos 2 \psi  \nonumber\\* &                      \qquad {} - \frac{D_0}{D} \cos\iota \sin2 \phi_0                                  \sin  2 \psi , \\
    \mathcal{A}^2 &= \frac{D_0}{D} \frac{\left(1 + \cos^2\iota\right)}{2}                          \cos2 \phi_0 \sin  2 \psi  \nonumber\\* &                      \qquad {} + \frac{D_0}{D} \cos\iota \sin2 \phi_0                                  \cos  2 \psi , \\
    \mathcal{A}^3 &= -\frac{D_0}{D} \frac{\left(1 + \cos^2\iota\right)}{2}                          \sin2 \phi_0 \cos  2 \psi  \nonumber\\* &                     \qquad {} - \frac{D_0}{D} \cos\iota \cos2 \phi_0                                  \sin  2 \psi , \\
    \mathcal{A}^4 &= -\frac{D_0}{D} \frac{\left(1 + \cos^2\iota\right)}{2}                          \sin2 \phi_0 \sin  2 \psi  \nonumber\\* &                     \qquad {} + \frac{D_0}{D} \cos\iota \cos2 \phi_0                                  \cos  2 \psi .
\end{align}
\end{subequations}
These terms are dependent on four parameters of the source:  its distance, $D$; 
the coalescence phase, $\phi_0$; the polarisation angle, $\psi$; 
and the inclination angle, $\iota$.  $D_0$ is a scaling distance, which is used in
normalizing the waveforms $h_{0, \pi/2}$.

The \ac{GW} signal seen by detector $X$ is a combination of the two polarisations 
weighted by the detector antenna response, $F^{X}_{\{+,\times\}}$~\cite{Jaranowski:1998qm},
\begin{equation}
    h^X(t) = F_+^X h_+(t^{X}) + F_\times^{X} h_\times(t^{X}).
\end{equation}
where the time of arrival in detector $X$ depends upon the sky location of the source relative
to the detector and the time of arrival at a fiducial location, for example the Earth's centre 
\cite{Harry:2010fr}.

The matched-filter is described by an inner product between a template \ac{GW} waveform of the above form, $h$, and the detector data $s$.
In general, the inner product between two such time series, $a^X$ and $b^X$, is given by
\begin{equation}
    \left(a^{X} | b^{X} \right) = 4 \operatorname{Re} \int_0^\infty                              \frac{\tilde{a}^{X}(f) \cdot [\tilde{b}^{X}(f)]^{\star}}{S^{X}_h(f)},
\end{equation}
where $S_h^X(f)$ is the noise \ac{PSD} in detector $X$.
For a network of detectors, we define the multi-detector inner product 
as the sum of the single detector inner products,
\begin{equation}
    (\mathbf{a | b}) \equiv \sum_{X=1}^{D} \left(a^{X} \middle| b^{X} \right) \, 
\end{equation}
where $D$ is the number of detectors in the network.
The multi-detector log-likelihood is then calculated as~\cite{Harry:2010fr},
\begin{align}
    \ln \Lambda &= (\mathbf{s | h}) - \frac{1}{2} (\mathbf{h | h}) \nonumber \\
                &= \mathcal{A}^\mu (\mathbf{s|h_\mu}) -                        \frac{1}{2}\mathcal{A}^\mu \mathcal{M}_{\mu \nu}                                  \mathcal{A}^\nu,
    \label{eq:loglike}
\end{align}
where $\mathbf{h_\mu} = (\mathbf{F_+ h_0, F_\times h_0, F_+ h_{\pi/2}, F_\times h_{\pi/2}})$, and
\begin{equation}
    \mathcal{M}_{\mu\nu} \equiv (\mathbf{h_\mu | h_\nu}).
\end{equation}
Maximising the log-likelihood over the values of $\mathcal{A}^{i}$, 
the coherent \ac{SNR} is defined as
\begin{equation}
    \rho^2_{\mathrm{coh}} \equiv 2 \ln \Lambda |_{max} =         (\mathbf{s | h_\mu}) \mathcal{M}^{\mu\nu} (\mathbf{s | h_\nu}),
    \label{eqn:snr}
\end{equation}
where $\mathcal{M}^{\mu \nu}$ is the inverse of $\mathcal{M}_{\mu \nu}$.

We can rewrite eqn (\ref{eqn:snr}) in a manner that makes it easier to compare to the coincident search.
To do so, we introduce the complex \ac{SNR} $z^{X}$ in detector $X$ as
\begin{equation}
z^{X} = (s^{X} | h^{X}_{0} ) + i (s^{X} | h^{X}_{\pi/2}) \, .
\end{equation}
Then, we can write the coherent \ac{SNR} as
\begin{equation}\label{eq:snr_proj}
\rho^2_{\mathrm{coh}} = \sum_{X, Y = 1}^{D} z^{X} P_{XY} z^{Y} \, ,
\end{equation}
where $P_{XY}$ is a projection of the SNR onto the 2-dimensional signal space:
\begin{equation}\label{eq:projection}
P_{XY} =  \left[
\frac{(\sigma^{X} F^{X}_{+}) (\sigma^{Y} F^{Y}_{+})}{\sum_{Z} (\sigma^{Z} F^{Z}_{+})^2}  
+ \frac{(\sigma^{X} F^{X}_{\times}) (\sigma^{Y} F^{Y}_{\times})}{\sum_{Z} (\sigma^{Z} F^{Z}_{\times})^2} 
\right] \, ,
\end{equation}
and $\sigma^{X} = \sqrt{(h_{0} | h_{0})_{X}} $ encodes the sensitivity of each detector.

Meanwhile the coincident \ac{SNR} is given by 
\begin{equation}
\rho^2_{\mathrm{coinc}} = \sum_{X, Y = 1}^{D} z^{X} \delta_{XY} z^{Y} \, .
\end{equation}
Thus, for a signal in the absence of noise (i.e. $P^{XY} z_{Y} = z^{X}$) the coincident and 
coherent \acp{SNR}
are identical.  For noise events, the coincident \ac{SNR} includes all of the noise, while
the coherent SNR incorporates only those contributions which are compatible with a 
coherent signal at all detectors.

\subsection{Covering the sky}
\label{sec:sky}

The coherent \ac{SNR} derived above depends upon the sky location of the source in two ways: 
through its dependence on the detector sensitivities, encoded in $F_{+,\times}$,  and through the 
differences in arrival time of the signal at the different detectors.
Consequently, the value of the coherent \ac{SNR} will change depending on the sky location of the 
source.  There is no analytic way to maximize over the sky position and instead we must search 
over a discrete grid of sky points, much as we do for the binary masses and spins.  The density of 
points required will depend upon the template's autocorrelation time \cite{Keppel:2013uma}.
For binary mergers, this depends upon the bandwidth of the signal \cite{Fairhurst:2009tc} which is 
typically around $100$ Hz and varies only slowly across the mass parameter space.  In what 
follows, we neglect the mass dependence of the sky grid and instead place a grid which is 
sufficiently dense for all templates.  While this results in a (somewhat) over-dense grid for higher 
mass templates, the effect is small.  

The effect of timing on the coherent \ac{SNR} is more significant than the change of the antenna 
response of the detectors \cite{Keppel:2013uma}.  In addition, the antenna 
responses change slowly in regions of the sky where the detectors have good sensitivity and most 
rapidly near the nulls of the detector.  This further reduces the significance of the changes in
$F_{+, \times}$ and consequently we ignore these effects and place points in the sky based solely on
time delays.  We do, however, make use of the appropriate detector responses when performing the 
search

\subsubsection{Sky tiling for a two-site network}
\label{sec:two_site_grids}

Sky coverage is significantly easier for a two-site network than one with three or more sites.  For such 
a network, there exists only a single time-delay baseline between the observatories meaning that we
can use a one-dimensional sky grid.  As we have remarked earlier, a coherent search of a 
two \textit{detector} network is no different from a coincidence search.  However, 
for the first eighteen months of LIGO's \glsreset{S5}\ac{S5} (November 2005 -- April 2007), 
the three \ac{LIGO} detectors formed a two-site, three-detector network, with both \ac{LHO}
instruments taking part in the run alongside a single instrument at \ac{LLO}.  In this case, 
there is a benefit to performing a coherent analysis.  This is an ideal first test case as we 
can explore the effects of a coherent search, but with a reduced 
sky grid. The second-generation \ac{aLIGO} instruments will form a two-detector, 
two-site network during the first observing run in 2015 \cite{TheLIGOScientific:2014jea, 
Aasi:2013wya}.  However, there is no good reason that a
coherent analysis would offer improved sensitivity to a coincidence one for this network.

For a two-site network, localisation by triangulation will reconstruct only the difference in time of arrival between the sites, mapping to a ring on the celestial sphere.
Furthermore, for the initial LIGO network, the maximized coherent SNR is completely indepedent of the values of $F_{+,\times}$.
As a result, the most efficient tiling of the sky sphere for a two-site `all-sky' search is given by the 
one-dimensional space of physically allowed time-delays.

The allowed time delays are bounded by the light travel time, $T$, between sites.
If we choose a temporal resolution $\delta t$, then the size of the grid is
\begin{equation}
    N = \left\lfloor \frac{T}{\delta t} \right\rfloor.
\end{equation}
We have found empirically that $\delta t = 0.5$ms is an appropriate
value.\footnote{The GW data are downsampled to 4096Hz when performing this analysis.  
Since it is more straightforward to shift by an integer number of samples, we actually use a time shift
of $2/4096$ seconds.}
The light travel time between \ac{LHO} and \ac{LLO} is $T = 10\,$ms so 
the two-site \ac{LIGO} grid requires 40 sky points. 

\begin{figure}
    \centering
    \includegraphics[width=.4\textwidth]{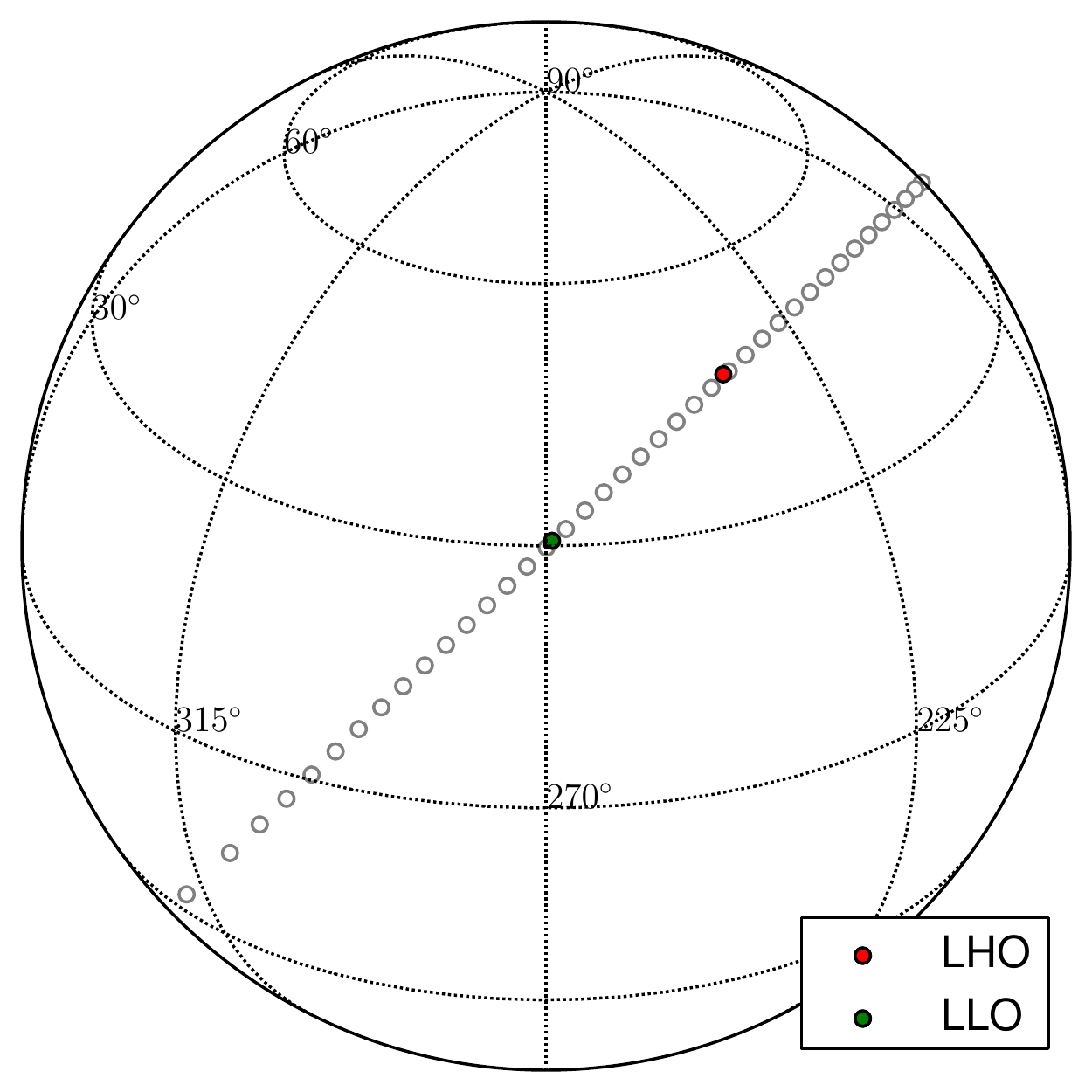} 
    \caption[A two-site all-sky grid for the LIGO detector network]            {A two-site all-sky grid for the \protect\ac{LIGO} detector              network.
             The points span the allowed time-delays between sites, forming              a great circle passing overhead both.}    \label{fig:two_site_grid}
\end{figure}

To seed this grid, we lay points with time delay $\tau$ in the range $[-T,T)$, evenly spaced by $\delta t$, and project these onto the prime meridian (zero longitude) via,
\begin{subequations}
    \begin{align}
        \phi &= 0\\
        \theta &= \cos^{-1}\left(\frac{\tau}{T}\right)
    \end{align}
\end{subequations}
This ring is then rotated onto the great circle containing both sites, as shown in 
\cref{fig:two_site_grid}. The grid  has greatest density orthogonal to the inter-detector baseline, 
where the time-delay is smallest, with density dropping symmetrically in either direction, as the 
time delay grows.

 \subsubsection{Sky tiling for a three-site network}
\label{sec:three_site_grids}

During the last six months of \ac{S5} \cite{Abadie:2010yb} and throughout \ac{S6} 
\cite{Colaboration:2011np}, the \ac{LIGO} and Virgo detectors 
operated a three-site network, allowing much more accurate time-delay triangulation, 
and better sky localisation~\cite{Cavalier:2006hi}.
This is likely to be the same network running during the middle years of the advanced detector era, 
after both \acl{aLIGO} and \acl{AdV} are observing, but before other detectors are operating.

With three detectors, the network can triangulate any signal to a single point in the 
hemisphere above the plane of the network.  A symmetry still exists in that plane, producing a 
second point in the other hemisphere. However, in many cases, the different detector responses
allow us to distinguish between these points \cite{VeitchMandel:2012}.
The unfortunate consequence of better sky localisation is the need for much larger sky grids 
for a full coherent analysis, increasing the computational cost of the search.

In order to map the sky for three sites, we follow the analytical models of \cite{Rabaste:2009mx}.
Consider a network of $D$ detectors, and define the time-delay vector
\begin{equation}
    \varmathbf{\tau} = \left( 
    \begin{array}{c}
    \tau_2 \\
    \vdots \\
     \tau_D
     \end{array}
     \right),
\end{equation}
where $\tau_n$ is the arrival time difference between detector $1$ and detector $n$.
Let $T_m$ be the light travel time between detector 1 and detector $m$, and 
define $\alpha_{mn}$ as the angle separating the lines connecting detectors 1 and $m$, and
detectors 1 and $n$.  Then, we can construct a bounding ellipse for the physically-admissible 
time delays,
\begin{equation}
    \varmathbf{\tau}^T\varmathbf{A}_D\varmathbf{\tau} \leq B_D,
\end{equation}
where, for the case of three detectors,
\begin{subequations}
\label{eq:timing_ellipse}
    \begin{align}
        \varmathbf{A}_3 &= \left[\begin{array}{cc}                               \frac{T_3^2}{T_2^2} &                                  -\frac{T_3}{T_2}\cos\left(\alpha_{23}\right)\\
                               -\frac{T_3}{T_2}\cos\left(\alpha_{23}\right) & 1
                             \end{array}\right]\label{eq:timing1}\\
        B_3 &= T_3^2\sin^2\left(\alpha_{23}\right).\label{eq:timing2}
    \end{align}
\end{subequations}
Here, $\cos \alpha_{23}$ effectively measures the correlation between the two time delays.
When $\cos \alpha_{23} = \pm 1$, the three sites lie in a line, the time delay matrix
$\mathbf{A}_{3}$ is singular and $B_{3} = 0$.  In this case, the time-delays are degenerate and 
the localization is no better than a two-site network.
When $\alpha_{23} = \pi/2$, the time-delays are independent ($\cos \alpha_{23} = 0$)
and the time-delay baselines form the major and minor axes of the bounding ellipse.
For values of $\alpha_{23}$ between these two extremes, there is some correlation between 
the time delays observed in the two detectors, and the ellipse of permitted time delays
will not be aligned with the baselines between detectors.

\begin{figure}
    \centering
    \includegraphics[width=.48\textwidth]{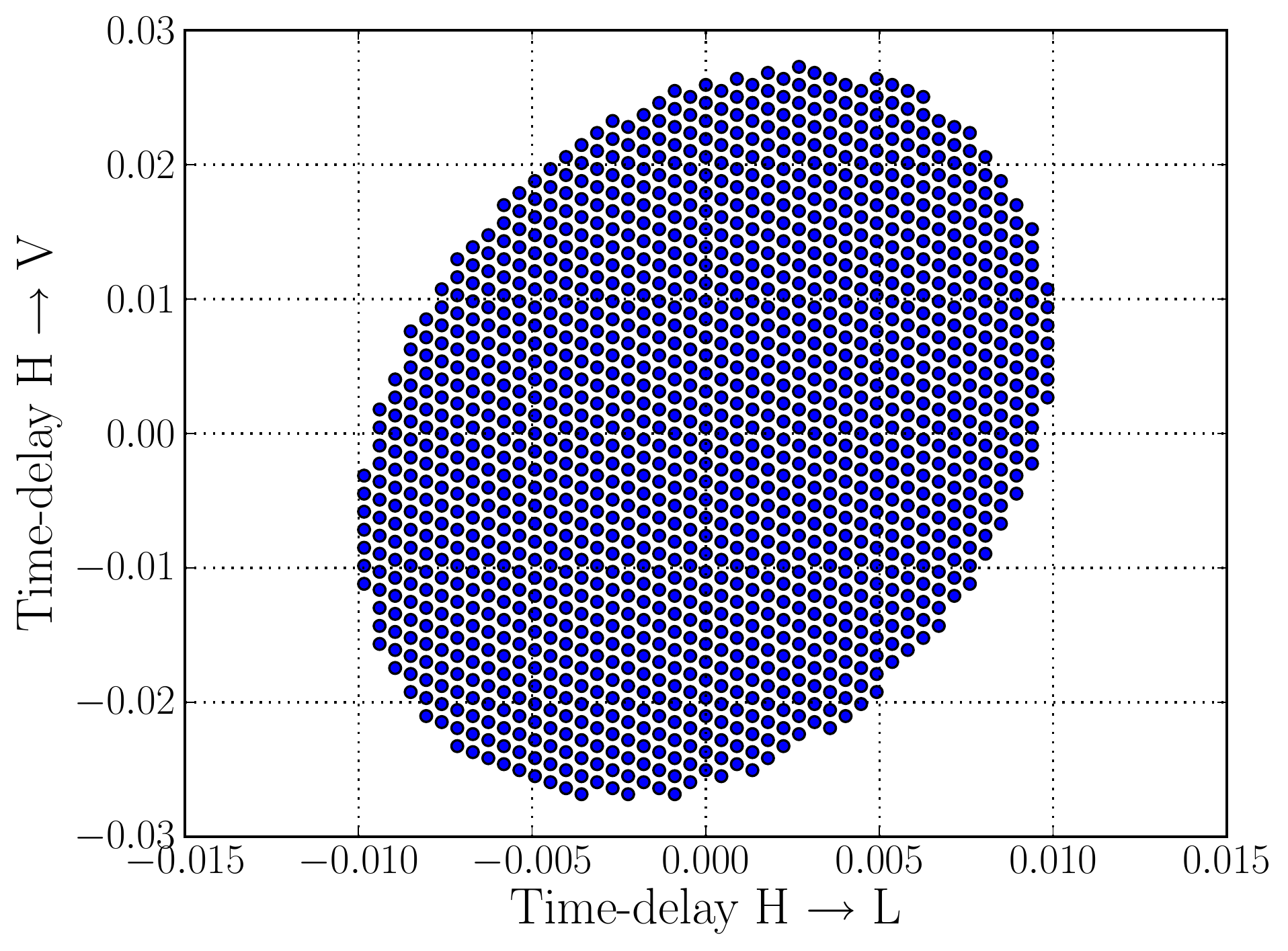}
    \caption[Time-delay tiles for the LIGO-Virgo three-site network]            {Time-delay tiles for the LIGO-Virgo three-site network.              All physically admissible points in this space are laid in              a hexagonal grid, with a minimal 0.5 ms spacing between              neighbouring points.}
    \label{fig:hlv_time_delay_grid}
\end{figure}

A grid of hexagonal time-delay tiles is laid in ($\tau_2$, $\tau_3$) coordinates, such that the distance between any two points matches the desired resolution.  In addition, all points must lie within the ellipse defined by \cref{eq:timing1,eq:timing2}.
The resulting time-delay grid for the three-site \ac{LIGO}-Virgo network is shown in \cref{fig:hlv_time_delay_grid}.

\begin{figure}
    \centering
    \includegraphics[width=.4\textwidth]{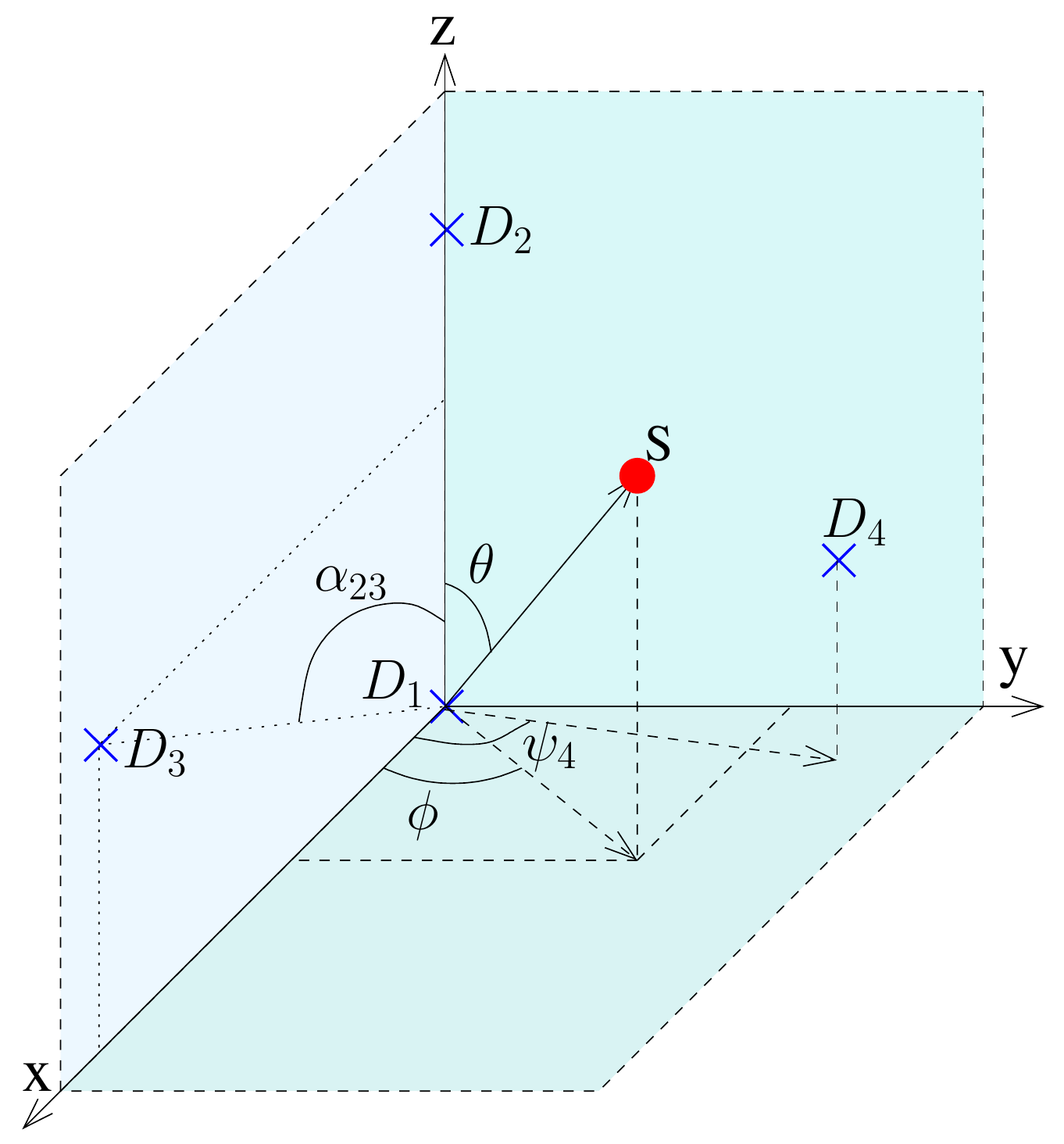}
    \caption[The network coordinate system used in projecting              points in time-delay space onto the sky.]            {The network coordinate system used in projecting
             points in time-delay space onto the sky~\cite{Rabaste:2009mx}.
             A three-site network defines a right-handed coordinate system,              with a potential fourth-detector breaking the symmetry $x$-$z$              plane.}
     \label{fig:rabaste_coordinates}
\end{figure}

The time-delay grid is then projected onto the celestial sphere by constructing a detector network coordinate system, as shown in \cref{fig:rabaste_coordinates}, where the time-delay coordinates are related to network longitude, $\phi$, and latitude $\theta$, via
\begin{subequations}
    \begin{align}
        \phi &= \pm \cos^{-1}\left(                  -\frac{T_2\tau_3 - T_3\tau_2\cos(\alpha_{23})}                        {T_3\sqrt{T_2^2 -\tau_2^2}\sin(\alpha_{23})}\right),\\
        \theta &= \cos^{-1}\left(-\frac{\tau_2}{T_2}\right).
    \end{align}
\end{subequations}
The network coordinates $(\phi, \theta)$ are then related to earth-fixed longitude and latitude with a simple rotation.
This projection is done twice, once for each hemisphere above and below the plane of the detector network.  In the end, we obtain a fixed grid in the Earth-centric frame.  

\begin{figure}
    \subfloat[]{
       \includegraphics[width=.23\textwidth]{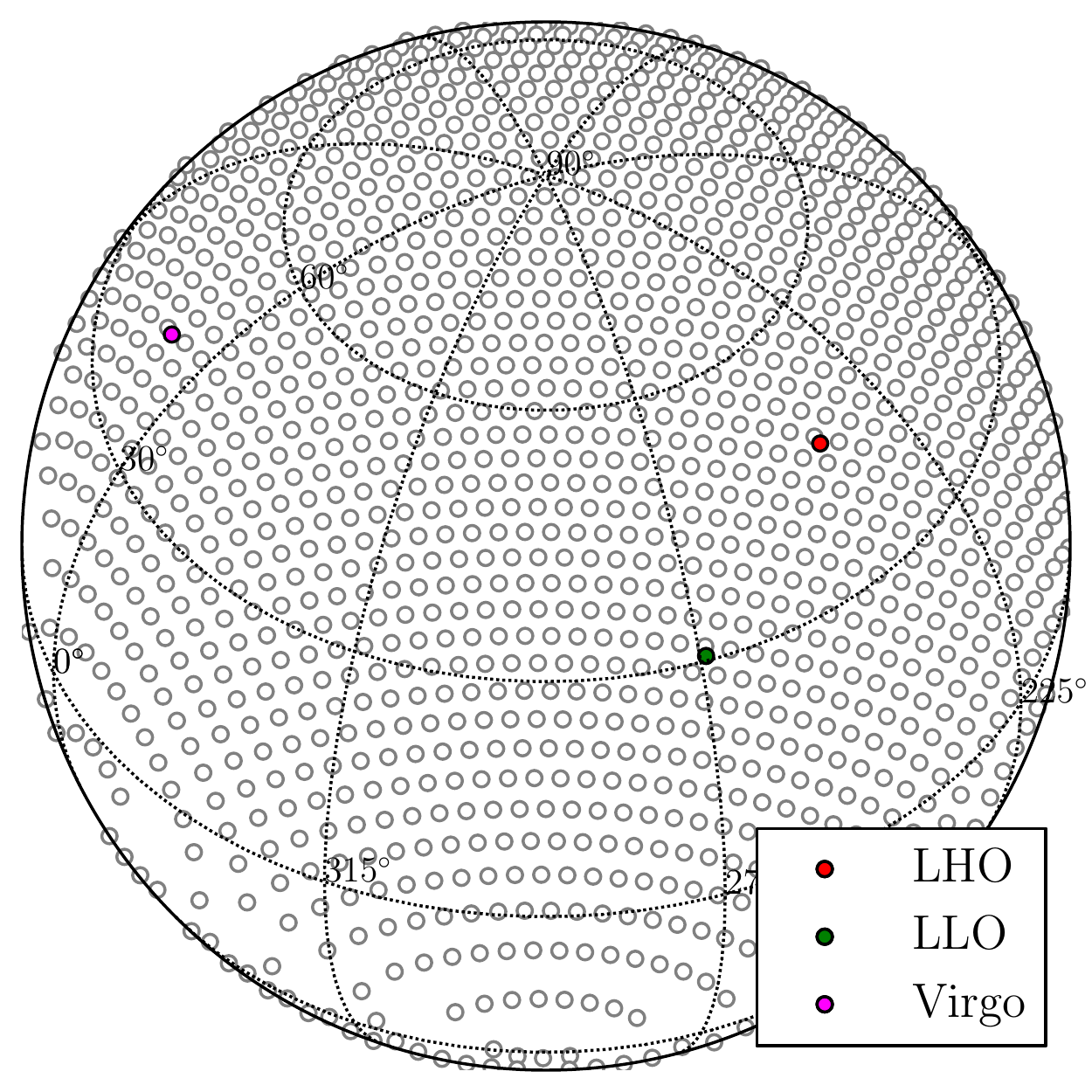}
       \label{subfig:three_site_overhead}
    }
    \subfloat[]{
       \includegraphics[width=.23\textwidth]{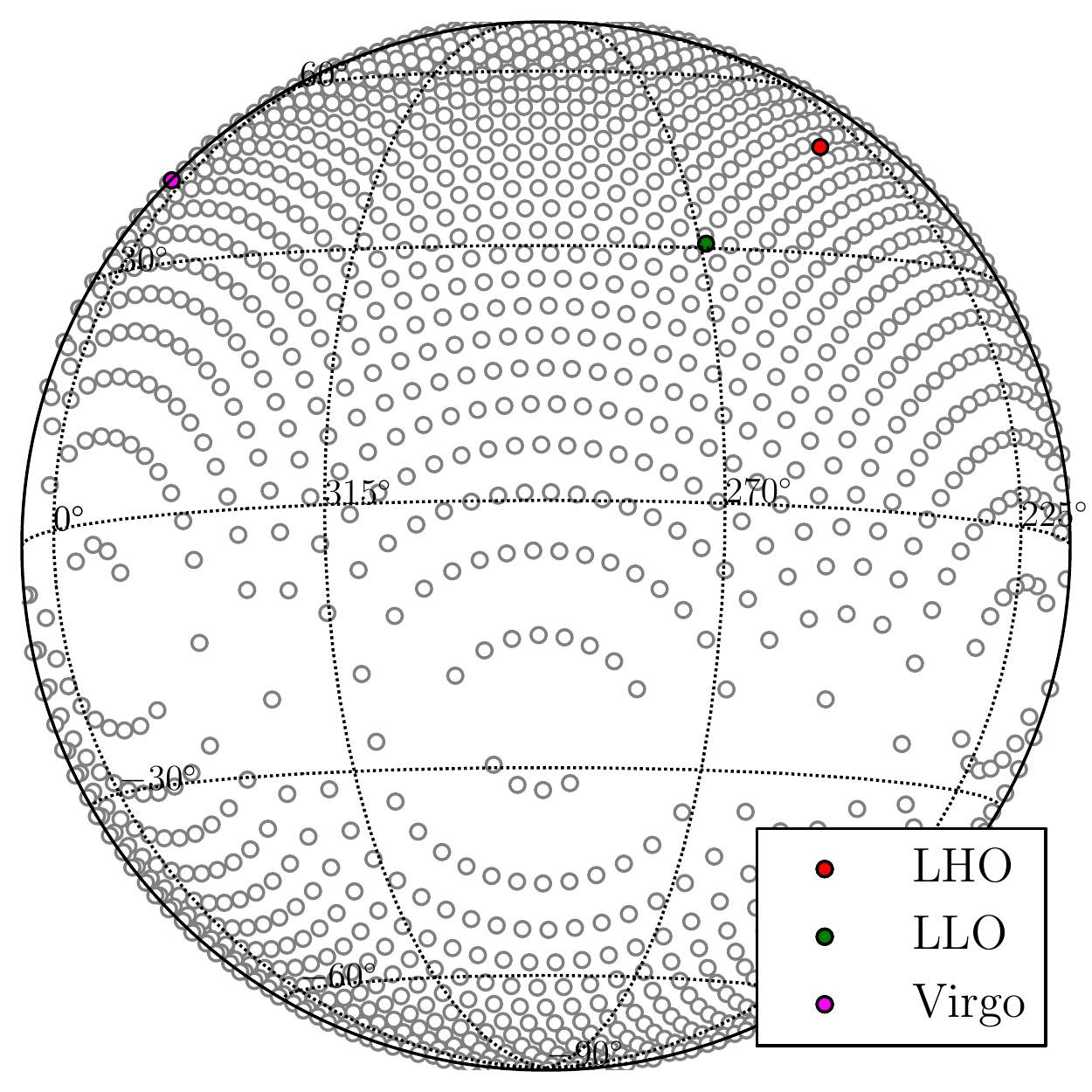}
       \label{subfig:three_site_plane}
    }
    \caption[The three-site all-sky coherent search sky grid for the              LIGO-Virgo detector network]
            {The three-site all-sky coherent search sky grid for the              \protect\ac{LIGO}-Virgo detector network.
             \protect\subref{subfig:three_site_overhead} and              \protect\subref{subfig:three_site_plane} view the same grid from              different angles.}
    \label{fig:three_site_grid}
\end{figure}

\Cref{fig:three_site_grid} shows an all-sky grid for the \ac{LIGO}-Virgo network; \ref{subfig:three_site_overhead} views the grid from nearly overhead the plane of the 
network, where the grid is densest, while \ref{subfig:three_site_plane} shows the relative 
sparsity of the grid parallel to the network plane.
This grid, using a time-delay resolution of $\delta t = 0.5$\,ms, contains over $2{,}700$ points,
representing a huge computational cost if applied na\"{i}vely to the coherent analysis.
 
\subsection{Searching over coalescence time}
\label{sec:time}

The matched filter between a template signal with coalescence time $t_{c}$ and the data
can be written as:
\begin{equation}
    \left(s | h_{t_{c}} \right) = 4 \operatorname{Re} \int_0^\infty                              \frac{\tilde{s}(f) \cdot [\tilde{h}_{t_{c} = 0}(f)]^{\star} e^{2\pi i f t_{c}}}{S_h(f)} \, .
\end{equation}                                                     
Therefore, it is possible to generate single detector \ac{SNR} time series efficiently by using
a Fourier transform \cite{Allen:2005fk}.  In practice, this is done by dividing the data into segments
of a fixed length and performing a \ac{FFT} on each segment.  Due to the
finite duration of the templates (and also the inverse power spectrum), 
filter wraparound will lead to the corruption of the SNR time series at the beginning and end
of each segment.  This effect is mitigated by simply overlapping the \ac{FFT} segments.

These single detector \ac{SNR} time series are then used to 
calculate the coherent \ac{SNR} as a function of time.  In practice, we find that it is not
necessary to calculate the coherent \ac{SNR} for every sky point and every time sample.  Instead,
we require that the single detector \acp{SNR} are above a threshold prior to proceeding with the
calculation of the coherent \ac{SNR}.  This greatly reduces the computational cost of the calculation
so that the computation of coherent SNR, for a three-site sky grid,
remains dominated by the calculation of the single detector \ac{SNR} time series.

\subsection{Searching over mass and spin}

The amplitude and phase evolution of the waveform depends sensitively on the masses and spins
of the binary components.  To search over the mass and (aligned-)spin parameters, we make
use of a discrete bank of template waveforms,
which covers the parameter space of binaries sufficiently densely that the difference between
any system and the nearest template is small enough that minimal signal power is lost 
\cite{Owen:1998dk, Cokelaer:2007kx}.

For the coincidence searches performed on the initial detector data, a separate template bank 
was constructed for each detector based upon its sensitivity \cite{Babak:2012zx}, where the density
of templates depends upon the noise power spectrum of the data from the detector.
For a coherent search, we must use the same bank for all detectors 
in the network.  However, at different points in the sky, the detectors have different antenna
responses and so contribute differently to the 
coherent analysis.  Thus, in principle, the template bank should be dependent upon the 
sky location.  However, as discussed above, for simplicity we do not do this.  Indeed, a detailed 
investigation \cite{Keppel:2013uma} showed that the
effect is minimal.  Instead, one would naturally use the harmonic mean of the detector PSDs to
construct an average PSD for generating the template bank \cite{Harry:2010fr}.  
For the analysis presented here, where the PSDs of the LIGO instruments have very similar shape,
we make a more straightforward choice and simply used a bank for one of the detectors as we found
that this had little effect on the results.

\subsection{Signal consistency tests and null SNR}
\label{sec:sig_consist}

In Gaussian data, the coincident or coherent \ac{SNR} would serve as a detection statistic: the
greater the value of the \ac{SNR} the less likely to arise due to noise fluctuations alone.  However,
in real detector data, there are numerous non-stationarities in the data which can lead to high
\ac{SNR} events.  Various techniques have been developed to mitigate the effect of these 
``glitches'' and get the search as close to the Gaussian limit as possible \cite{Allen:2005kaa, 
Hanna:thesis, Harry:2010fr, Babak:2012zx}.

\subsubsection{Null SNR}

A coherent gravitational wave search involves combining the data from the detectors in the network
to produce data streams that are sensitive to the two polarizations of gravitational radiation.  
When there are more than two detectors in the network, it is possible to
construct additional data streams which do not contain any gravitational wave contribution 
\cite{GuerselTinto1989}.  
Using the framework above, we denote the null \ac{SNR} as
\begin{equation}\label{eq:null_snr}
\rho^{2}_{\mathrm{null}} = \rho_{\mathrm{coinc}}^{2} - \rho_{\mathrm{coh}}^{2} \, .
\end{equation}
In Gaussian noise, this would be $\chi^{2}$-distributed with $2(d-2)$ degrees of freedom.
Removing this noise contribution from the coherent \ac{SNR} increases the sensitivity of the coherent search.
When the data contain non-stationary transients, they will tend to be observed in a single detector and will not be consistent with a coherent signal.  They will, therefore, have a large null \ac{SNR}, and we can remove events with high null \ac{SNR} from the search results.

\subsubsection{Single detector SNR threshold}

We also make use of thresholds on the single detector \acp{SNR} to reject events which are
unlikely to be caused by real signals.  By requiring that the \ac{SNR} is above a given threshold
in at least two detectors, we can eliminate the vast majority of events caused by non-stationary
transients, as they appear in only a single detector.  As discussed above, an additional benefit 
of the single detector thresholds is that they are very cheap to compute.  Consequently, we 
apply them before calculating the coherent \ac{SNR} in order to reduce the computational 
cost of the analysis.

\subsubsection{$\chi^{2}$ tests}

Finally, we make use of signal consistency tests, typically called $\chi^{2}$ vetoes  
\cite{Allen:2005kaa, Hanna:thesis}.  The basic
concept is to subtract the template that matches the observed signal and then check that what
remains is consistent with Gaussian noise.  This is done by filtering the residual data
with test waveforms, $T^{i}$, that are orthogonal to the best fit template and calculating the sum of 
squares of
\acp{SNR} in those templates.  In Gaussian noise, the value will be $\chi^{2}$-distributed with 4N
degrees of freedom (where N is the number of test waveforms used).  Any mis-match 
between the signal and the template will lead to imperfect cancellation of the signal
and a contribution to the $\chi^{2}$.  This mis-match could be due to the discreteness of the 
template-bank, differences between the true waveforms and those used as templates, or errors due to the calibration of the detectors.
To account for this, an \ac{SNR}-dependent threshold is
typically used when rejecting events with a large $\chi^{2}$ value.

In this analysis, we make use of three different $\chi^{2}$ tests, as implemented for the targeted
coherent search \cite{Harry:2010fr}:
\begin{enumerate}
\item \textit{Frequency bins:} The test waveforms $T^{i}$ are generated by chopping up the 
template $h(t)$ into $(N+1)$ sub-templates in the frequency domain, each of which contains an 
equal fraction of the power.  From these, we can construct $N$ orthonormal waveforms, all of 
which are orthogonal to $h(t)$.

\item \textit{Template bank:} The test waveforms $T^{i}$ are binary merger waveforms with
different mass and spin parameters.  In general, they will not be exactly orthogonal to $h(t)$,
so we simply remove the part proportional to $h(t)$.  Note, however, that the test waveforms
will also not be orthonormal, and consequently the expected distribution is not $\chi^{2}$-distributed with $4N$ degrees of freedom.  We do not attempt to resolve this issue but instead
apply an empirical threshold.

\item \textit{Autocorrelation:} The test waveforms $T^{i}$ are time-shifted copies of the original
waveform $h(t)$.  As with the template bank test, these waveforms are neither orthonormal nor
orthogonal to $h(t)$.  We proceed as above.

\end{enumerate}

 \section{Search implementation and testing}
\label{sec:two_site_search}
In this section we demonstrate the first implementation of a fully-coherent, all-sky search for signals from \acl{BNS} inspirals.
The full analysis calculates the coherent \ac{SNR} for each template in a bank, along with a number of signal-consistency statistics that allow distinction between non-Gaussian noise artefacts and true signals~\cite{Harry:2010fr,Williamson:2014}.

\subsection{Data selection}
\begin{figure}
    \centering
    \includegraphics[width=.55\textwidth]{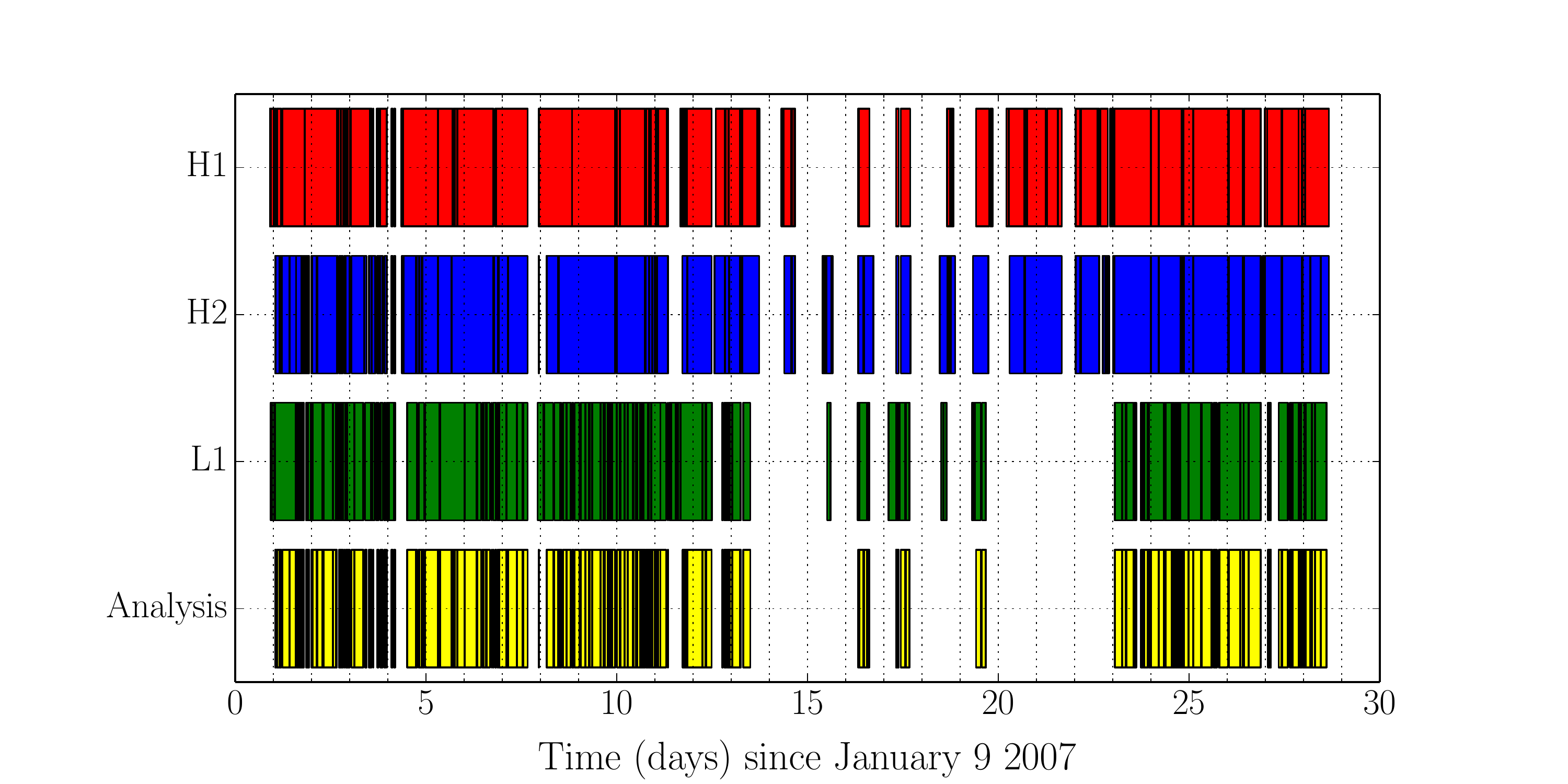}
    \caption[Analysis segments for coherent all-sky search of one month of              S5 data]
            {Analysis segments for coherent all-sky search of one month of              \protect\ac{S5} data.}
    \label{fig:all_sky_segments}
\end{figure}

\begin{figure}
    \centering
    \includegraphics[width=.48\textwidth]{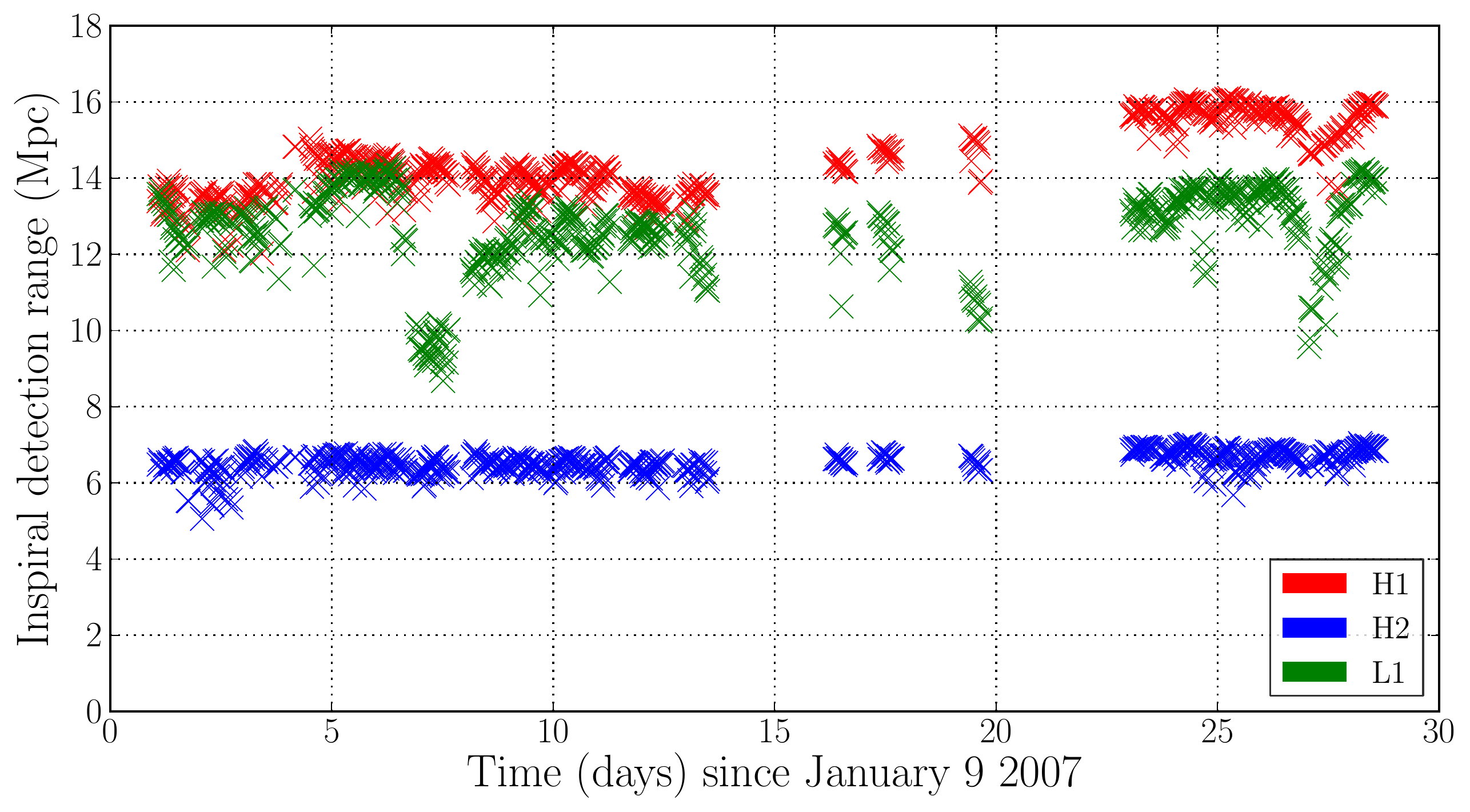}
    \caption[Inspiral averaged sensitive range for the LIGO network              during one month of S5]            {Inspiral averaged sensitive range for the \protect\ac{LIGO}              network during one month of \protect\ac{S5}.}
    \label{fig:all_sky_range}
\end{figure}

The coherent all-sky analysis was used to search one month of data from \ac{S5}, during which the 
three \ac{LIGO} instruments formed a two-site, three detector network \cite{Abbott:2007kv}.  As
usual, the 4km detectors at Hanford and Livingston will be denoted H1 and L1 respectively,
while the 2km Hanford detector is denoted H2.
Only those data segments during which all three detectors were operating nominally are used, with 
the additional requirement that each segment was longer than 2176 seconds to allow for accurate 
measurement of the detector noise \ac{PSD}; \cref{fig:all_sky_segments} shows the segments that 
were selected. The sensitivity of the detectors to neutron star mergers is characterized by the
sky and source orientation averaged distance at which the signal from a $1.4-1.4 M_{\odot}$ binary 
 would be observed with a
\ac{SNR} of 8 in a single detector.  \Cref{fig:all_sky_range} shows the sensitive range for each 
detector in the network during this period.
The smaller H2 detector maintained a range between 6--7\,Mpc throughout, while the larger 
instruments improved as the run progressed, with H1 peaking at 16\,Mpc.

\subsection{Generating a template bank}

\begin{figure}
    \centering
    \includegraphics[width=.48\textwidth]{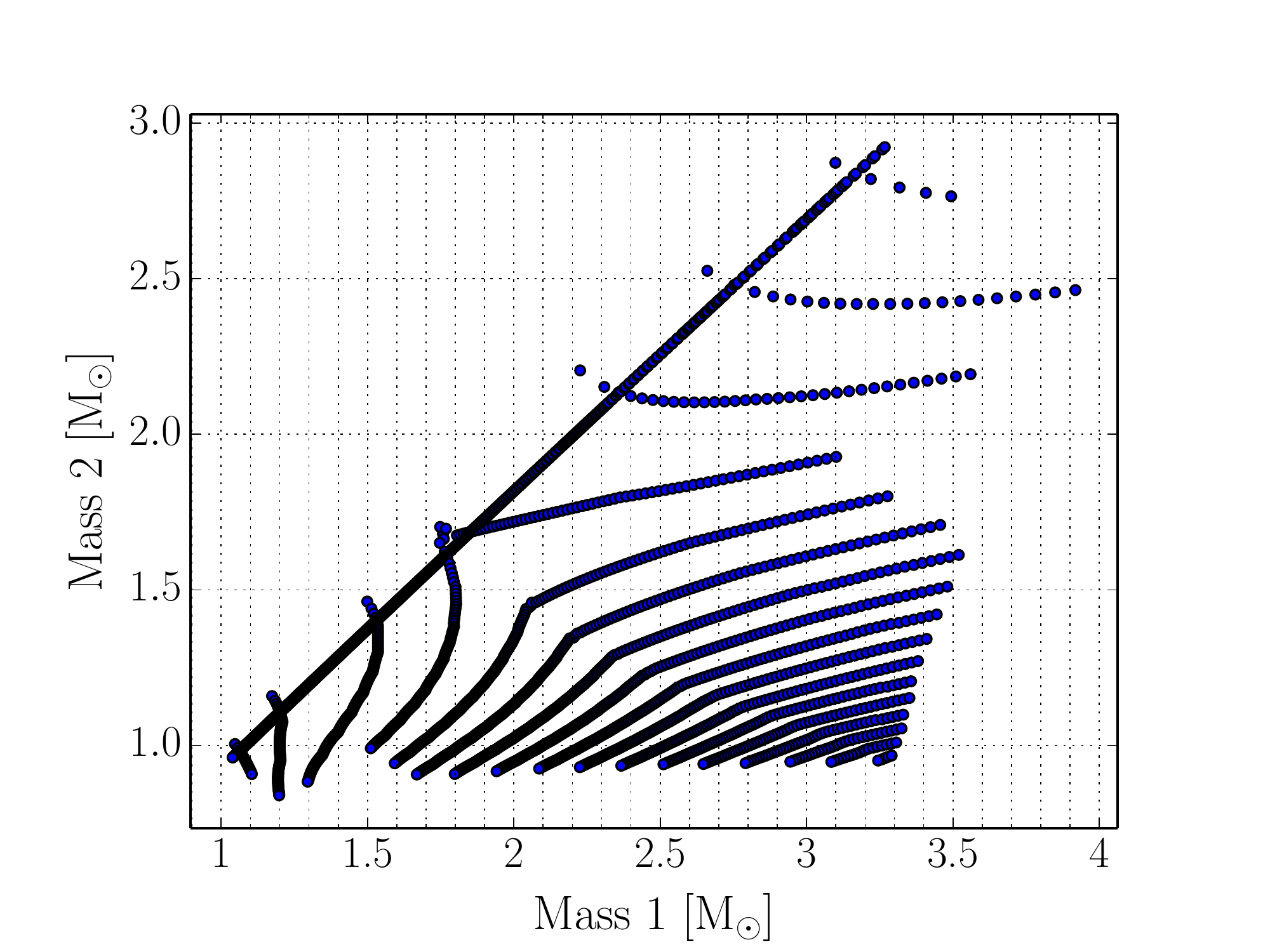}
    \caption[The template bank used for the coherent all-sky search of one              month of S5 data]
            {The template bank used for the coherent all-sky search of one              month of S5 data.}
    \label{fig:all_sky_tmpltbank}
\end{figure}

The coherent search was performed over the parameter space of neutron star binaries.  We take
a mass range of $1-3 M_{\odot}$ for the binary components, and neglect the effect of spins on
the waveform \cite{Brown:2012qf}.
A bank of template inspiral waveforms was laid using the methods of \cite{Cokelaer:2007kx,Harry:2010fr}, with a single-detector bank for the L1 detector used as a simple approximation to a fully-coherent template bank.
For the analysis, non-spinning, 3.5 \ac{PN}-order, \ac{BNS} inspiral waveforms were placed with a maximum combined mass of 6.2\,\msun, resulting in ${\sim}2{,}200$ templates, as shown in \cref{fig:all_sky_tmpltbank}.

\subsection{Performing the coherent analysis}
\label{sec:coh_analysis}

For computational ease, each analysis segment is divided into chunks of 2176 seconds, 
overlapping by 64 seconds on each end; a template bank is generated for each of these chunks
 (due to changing detector sensitivity), with each chunk processed separately.
The data from each detector are used to estimate its \ac{PSD}, using the average of sixteen 
50\% overlapped 256-second sensitivity estimates.

For each of the 256-second blocks, the data from each detector are filtered against each 
template in turn, producing time-series of single-detector \ac{SNR}.  The \ac{SNR} time series
at the beginning and end of each block is corrupted due to filter wraparound \cite{Allen:2005fk}, 
and only the middle 128-seconds is retained.
When the single detector \acp{SNR} are above threshold, the sky grid is used to construct 
coherent combinations of these time-series and calculate the coherent \ac{SNR}.  At any
time sample where the coherent \ac{SNR} for a given sky point is above threshold, the value
is recorded and the signal consistency tests calculated.

\subsubsection{Background estimation with time-slides}
\label{subsec:all_sky_slides}

The noise background is measured using time-shifted data.
Since a fully-coherent search combines detector data at the filtering stage, each time-slide 
analysis requires re-computing the coherent matched-filter \ac{SNR}.
This represents a huge computational cost, in comparison to the coincidence-based analysis 
where time-shifts are performed on the single-detector events produced after filtering.

In this analysis, 10 time-shifts were constructed, each sliding data from the L1 detector forward 
by a multiple of 128-seconds.  Such large slides are computationally simple and performed by 
combining data from the \ac{LHO} instruments with those from L1 in a different 256-second block.
The slides are performed on a ring (formed by the sixteen analysis blocks for a single chunk), 
whereby any L1 data slid off the end of the analysis chunk is re-inserted at the start and filtered 
against the data from the \ac{LHO} instruments in the first block.  The computational cost for each
of these time shifts is equal to the original analysis.

\subsubsection{Simulations}
\label{subsec:all_sky_injections}

A set of non-spinning \ac{BNS} inspiral simulations were used to inform tuning of the signal-consistency cuts and the detection statistic, and measure search performance through simulation recovery efficiency.
The signals were uniformly distributed in mass (with component masses
between $1 M_{\odot}$ and $3 M_{\odot}$), sky location and orientation.  Observed signals are
expected to be distributed uniformly in volume.  However, if simulations are distributed uniformly
in volume, the vast majority of simulated signals will be at large distances and below the
detection threshold of the pipeline.  Instead, we generate simulations with distances uniformly 
distributed between $1-60$\,Mpc.  

Throughout the following descriptions of signal-based and \acl{DQ} cuts, we use time-shifted
and simulated events to assess the sensitivity of the analysis, and to determine appropriate
thresholds to separate signal from background.

\subsection{Event down-selection}

While the data from \glsuseri{GW} detectors are often modelled as Gaussian, in practice this is 
rarely the case.
The data are regularly contaminated with non-stationary, and non-Gaussian noise artefacts 
(`glitches') that will be detected, even in a coherent analysis, with high \ac{SNR}.
As a result, the rate of events identified as significant by the matched-filter is too high to be either
useful or practical -- large noise glitches will trigger across the full template bank, producing multiple 
events from a single noise input.

Events are down-selected by identifying those most significant relative to the surrounding data.
The full list of events are divided into 100\,ms bins, with an event selected only if it is the loudest 
in its own bin, and louder than all events in a 100\,ms window around itself.
This selection method typically reduces the event rate by a factor of 100 or more, by identifying those event triggers that represent the peak of an excess power transient in the data (either noise or \ac{GW} signal).

\subsubsection{Signal-consistency cuts}

\begin{figure*}
    \centering
    \subfloat[]{        \includegraphics[width=.48\textwidth]{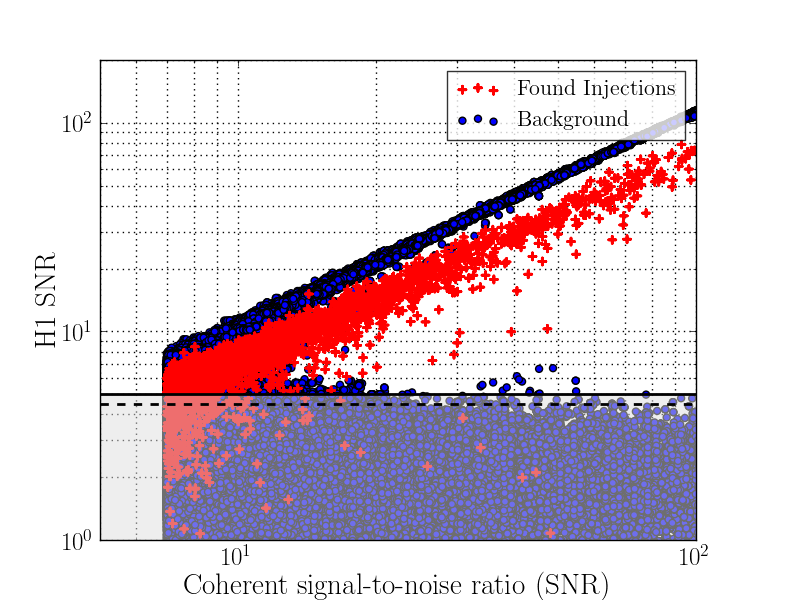}
        \label{subfig:all_sky_h1_snr}}
    \subfloat[]{        \includegraphics[width=.48\textwidth]{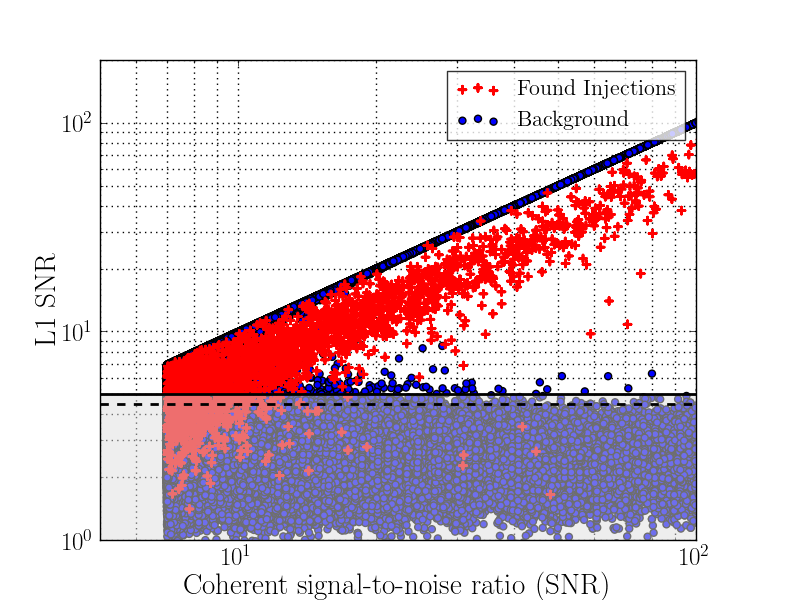}
        \label{subfig:all_sky_l1_snr}}
    \caption[The impact of the single-detector SNR cut on events from an              all-sky coherent search]            {The impact of the single-detector \protect\ac{SNR} cut on              events from an all-sky coherent search.              The blue dots are those from the noise background, while the              red pluses are those from simulated \protect\ac{BNS} signals.
             The shaded region represents the single-detector cut as applied.
             All events with power only in a single-detector are vetoed as              likely noise artefacts.}
    \label{fig:all_sky_sngl_snr}
\end{figure*}

Each of the signal-consistency tests outlined in \Cref{sec:sig_consist} are applied 
equally to the event triggers 
from the foreground data, each of the time-slide background trials, and the simulations.
The \ac{SNR} thresholds for the search are:
\begin{enumerate}[i]
    \item single-detector \ac{SNR} $\ge 5$ in one detector,           and $\ge 4.5$ in a second detector,
    \item coherent \ac{SNR} $\ge 7$.
\end{enumerate}

The first signal-consistency cut, on single-detector \ac{SNR}, identifies those events with power
in a single detector only, typically removing more than 90\% of all events.
\Cref{fig:all_sky_sngl_snr} shows the impact of this cut on the events from the S5 analysis, applied 
to both H1 and L1 data, including both the background from time slides and from the simulation set.
Those background \glspl{trigger} (blue dots) on the diagonal in each figure represent events with 
power only in that detector, and fail the cut (black line) in the other detector.
The simulated signals are below the diagonal because their coherent \ac{SNR} is accumulated 
from a strong component in each detector.

The $\chi^{2}$-based signal consistency tests are used to further reduce the impact of noise 
triggers.  To do so, we first calculate the re-weighted \ac{SNR}~\cite{Babak:2012zx}, 
in a similar way as was done for the coincidence analysis, as:
\begin{equation}\label{eq:new_snr}
    \rho_{\mathrm{\chisq}} =         \left\{\begin{array}{cl}            \displaystyle \rho, & \chisq \le n\sub{dof} \\
            [3mm]
            \displaystyle \frac{\rho}{\left[\left(1 +                                        \left(\frac{\chisq}{n\sub{dof}}\right)                                           ^{3}\right)/2\right]^{1/6}},                & \chisq > n\sub{dof},
        \end{array}\right.
\end{equation}
where $n\sub{dof}$ denotes the number of degrees of freedom for the $\chi^{2}$ tests.

\begin{figure*}
    \centering
    \subfloat[]{        \includegraphics[width=.48\textwidth]{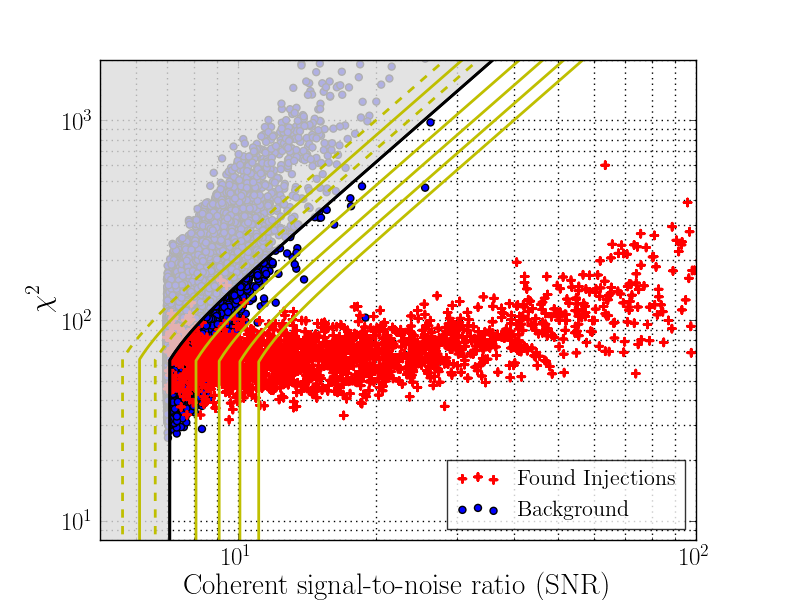}
        \label{subfig:all_sky_chisq}}
    \subfloat[]{        \includegraphics[width=.48\textwidth]{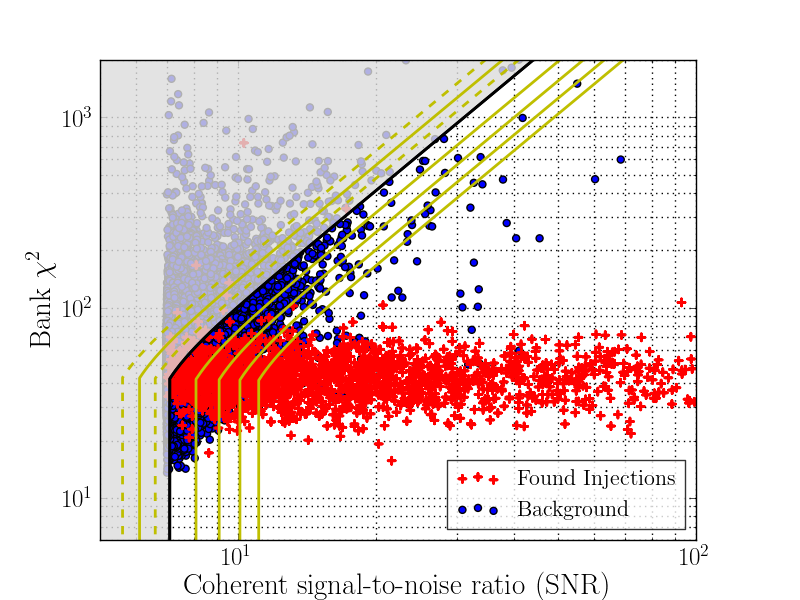}
        \label{subfig:all_sky_bank_chisq}}\\
    \subfloat[]{        \includegraphics[width=.48\textwidth]{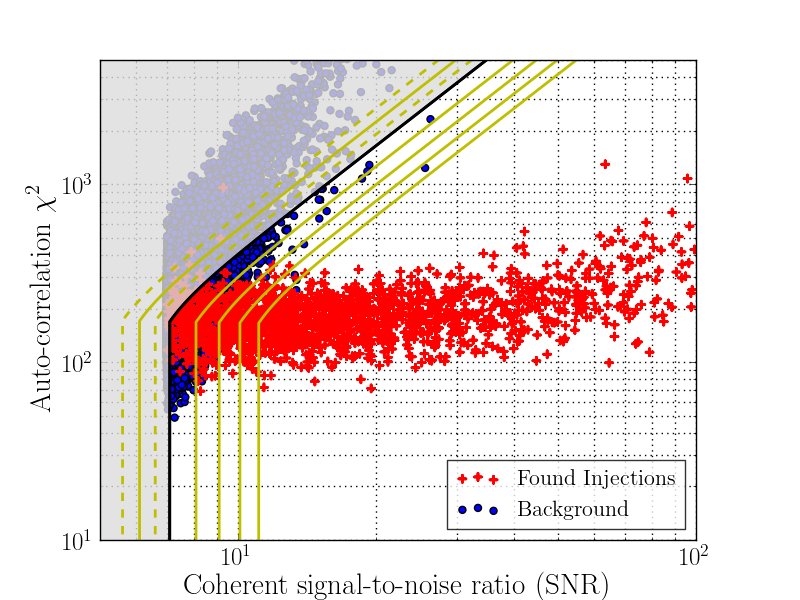}
        \label{subfig:all_sky_auto_chisq}}
    \subfloat[]{        \includegraphics[width=.48\textwidth]{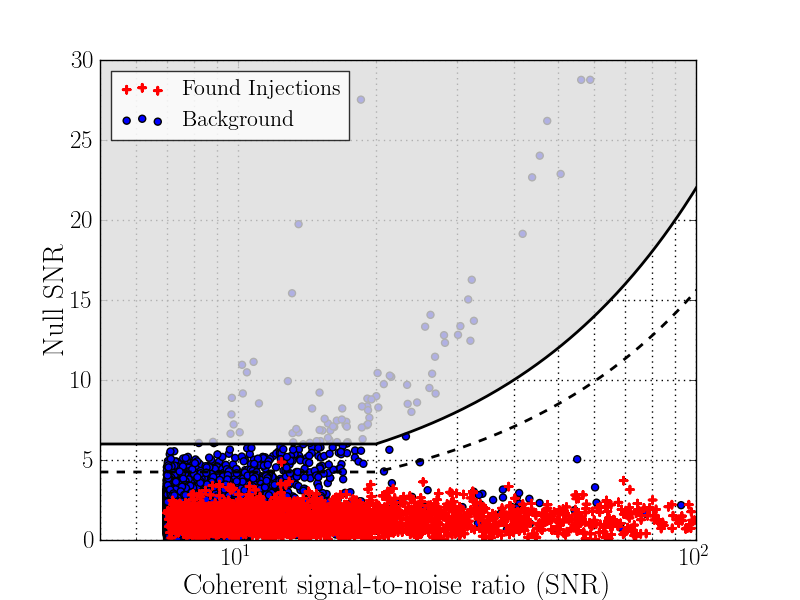}
        \label{subfig:all_sky_null}}
    \caption[The impact of signal-consistency cuts on the background of an              all-sky coherent search]            {The impact of signal-consistency cuts on the background of an              all-sky coherent search.
             The blue dots are background events from time-slid analysis,              and the red pluses are events from \protect\ac{BNS} simulations,
             and the shaded region covers those events failing              the signal-consistency test.
             In figures \protect\subref{subfig:all_sky_chisq} to              \protect\subref{subfig:all_sky_auto_chisq}, the contours              represent constant values of the re-weighted \protect\ac{SNR}              (dashed lines for half-integers, solid for integers).              The detection statistic is constructed from the \chisq              re-weighted \protect\ac{SNR}, and the null \protect\ac{SNR}, 
             figure \protect\subref{subfig:all_sky_null}, for which events              above the dashed line are down-ranked using              \protect\cref{eqn:snr_det}.}
    \label{fig:all_sky_sbv}
\end{figure*}

We then apply the following thresholds:
\begin{enumerate}[i]
  \setcounter{enumi}{2}
    \item frequency-bin \chisq re-weighted \ac{SNR} $> 7$,
    \item template-bank \chisq re-weighted \ac{SNR} $\ge 7$,
    \item autocorrelation \chisq re-weighted \ac{SNR} $\ge 7$.
    \item an SNR dependent threshold on the null SNR:
    \begin{equation}
    \rho_{\mathrm{null}} \le \left\{\begin{array}{cl}            \displaystyle 6, & \rho \le 20,  \\
            [3mm]
            \displaystyle 6 + \left(\frac{\rho - 20}{5}\right) \,  & \rho > 20,
        \end{array}\right.
\end{equation}
\end{enumerate}
\Cref{subfig:all_sky_chisq,subfig:all_sky_bank_chisq,subfig:all_sky_auto_chisq} show the impact 
of the three \chisq consistency cuts, each evaluated after the single-detector \ac{SNR} cut has 
been applied.
These statistics clearly differentiate between the recovered simulations and the noise 
background, removing those events inconsistent with a true signal.
\Cref{subfig:all_sky_null} shows the impact of the null \ac{SNR} cut, similarly evaluated after the single-detector \ac{SNR} cut has been applied.
The null \ac{SNR} cuts a relatively small number of noise events that are incoherent between detectors.

The final detection statistic is constructed using two signal-consistency statistics to 
down-rank likely noise events.  First, we use the frequency-bin $\chi^{2}$ test to re-weight
the \ac{SNR}, using \cref{eq:new_snr}.  The curves in figure \cref{subfig:all_sky_chisq} are
lines of constant $\rho_{\chi^{2}}$ and show how high \ac{SNR} noise events are down-weighted
while simulated signals have a re-weighted \ac{SNR} similar to their coherent \ac{SNR}.
Second, we use the null \ac{SNR} to further re-weight \ac{SNR} into the detection statistic.
We introduce an \ac{SNR} dependent threshold, $\xi_{\mathrm{null}}$:
\begin{equation}
    \xi\sub{null} =  \left\{\begin{array}{cl}
                            4.25, & \rho \le 20 \\[0.1in]
              \displaystyle 4.25 + \left(\frac{\rho - 20}{5}\right), & \rho > 20
                            \end{array}\right. 
    \label{eqn:null_thresh_def}
\end{equation}
and down-weight any triggers for which the null \ac{SNR} is greater than this value.
The dashed line in figure \cref{subfig:all_sky_null} shows the threshold above which events
are down-weighted based upon their null \ac{SNR}.  This downweighting reduces the significance
of many background events, but affects only a handful of simulated signals.

The final detection statistic is
\begin{equation}
    \rho\sub{det} =  \left\{\begin{array}{cl}
                            \rho_{\chisq}, & \rho\sub{null} \le \xi\sub{null} \\[0.1in]
              \displaystyle \frac{\rho_{\chisq}}{\rho\sub{null}- (\xi\sub{null} - 1) },                                & \rho\sub{null} > \xi\sub{null} \, .
                            \end{array}\right.
    \label{eqn:snr_det}
\end{equation}

\subsubsection{Data quality cuts}

Instrumental and environmental disturbances can lead to periods of poor data quality in the
detectors.  The signal consistency tests described above are used to mitigate the impact of these
poor data.  In addition, \ac{DQ} vetoes are also used to identify noise artefacts in the data, using 
instrumental and environmental correlations.  We use the same data quality definitions
as used in the coincidence search \cite{Abbott:2009qj} (for a more detailed description of how
they are utilized, see \cite{Abbott:2009tt, Babak:2012zx}).
Three different categories of data quality are generated. 
Data that are too poor to use at all, and would corrupt the \ac{PSD} estimate are labelled \textit{category 1}.
When the impact is less severe, it is preferable to include the data in the analysis, and then 
remove any triggers at times of poor data quality. If we were to remove these data prior to filtering, 
we would lose an additional 64s on either side as discussed in \Cref{sec:coh_analysis}.
\textit{Category 2} data quality vetoes identify times of instrumental problems with known correlations to the \ac{GW} channel, while \textit{category 3} vetoes identify times of poor data, often identified by statistical
correlations with the \ac{GW} channel.  

\begin{figure}
    \centering
    \includegraphics[width=.48\textwidth]{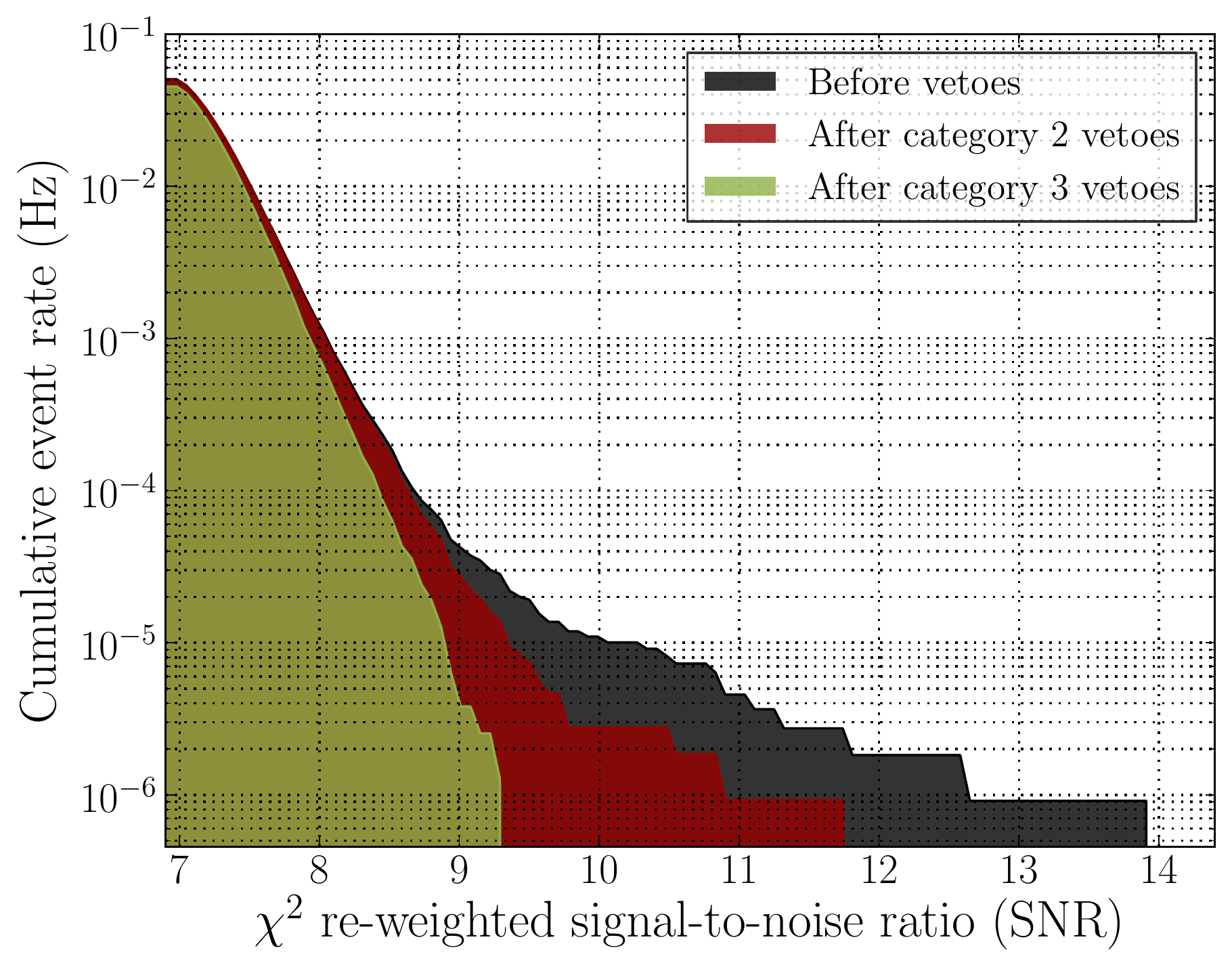}
    \caption[The impact of data quality vetoes on the background of an              all-sky coherent search]            {The impact of data quality vetoes on the background of an              all-sky coherent search. As seen, the loudest event is reduced              from a re-weighted \protect\ac{SNR} of 13.9 to 9.3.}    \label{fig:all_sky_vetoes}
\end{figure}

These \ac{DQ} vetoes are applied to data from each of the foreground, background and simulations 
such that if an event is vetoed in any one instrument then it is removed from the search~\cite{Slutsky:2010jn}.
\Cref{fig:all_sky_vetoes} shows the impact of category 2 and 3 \ac{DQ} vetoes on the background 
events (from time slides), after the application of the signal-based vetoes.
The category 2 vetoes are successful in removing the very loudest events, with the loudest event 
reduced from 13.9 to 11.8.  In applying the category 2 vetoes, we discard around 1\% of the
available data.  Category 3 vetoes are effective at removing the remaining tail of loud
events in the coherent search, with the loudest event reduced to a detection statistic value of 9.3. 
However, to achieve this, we lose 25\% of the available data since a \ac{DQ} veto in any one of the
three detectors leads to the data being discarded from the coherent analysis.  One may question whether there is benefit in applying the vetoes if they lead to the removal of such a large amount of
data.  We have removed 25\% of the data and succeeded in reducing the \ac{SNR} of the
loudest background events by a similar amount.  This equates to improving the distance
reach of the search by the same amount.  Since sources are expected to be uniformly
distributed in time and \textit{volume}, imposing \ac{DQ} cuts has improved the sensitivity
of the search by around 50\%.

The amount of time removed due to the data quality vetoes is still too high to be acceptable in 
a real search.  However, for much of this time, the data quality is poor in only one detector.  
Therefore, it is possible to ``recover'' this time by performing a two-detector
analysis on this data.  Indeed, this is the procedure that has been followed in the coincidence
search \cite{Abbott:2009qj, Abbott:2009tt, Babak:2012zx}.
As we have noted previously, there is no reason for a two detector
analysis to be performed coherently, so it would be natural to run the coincidence search
over these times.  

We can qualify the overall impact of the combined signal-based and data-quality vetoes, and the effectiveness of the chosen detection statistic, by comparing the high coherent \acp{SNR} seen in \cref{fig:all_sky_sngl_snr} to the final distribution in \cref{fig:all_sky_vetoes}.

\section{Search performance}
\label{sec:all_sky_efficiency}

The performance of the search is measured using the results of the simulation run, after all signal
consistency tests and \acl{DQ} cuts have been applied.
All simulations for which no event was recorded are classed as missed.
Those simulations with an associated trigger with a larger value of the detection statistic than all of the background events are classed as recovered, while those events with an associated trigger that is not louder than all background events are `marginally' recovered.

\begin{figure}
    \centering
    \includegraphics[width=.48\textwidth]{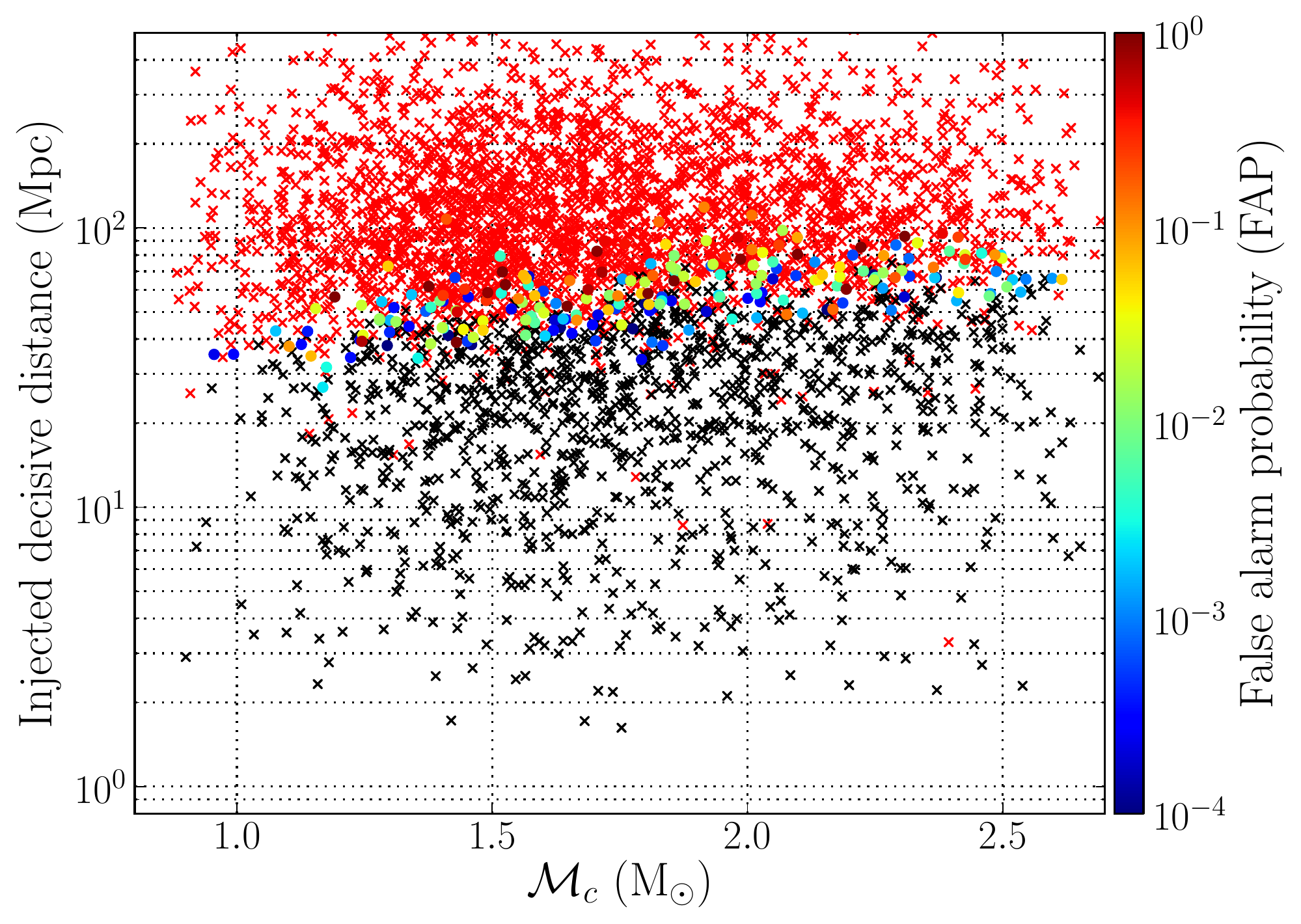}
    \caption[Recovery of simulated BNS signals during the S5 coherent              all-sky analysis]            {Recovery of simulated \protect\ac{BNS} signals during the              \protect\ac{S5} coherent all-sky analysis.              Successfully recovered signals as marked with black crosses,
             missed signals with red crosses, and marginally-recovered              signals with dots coloured by false alarm probability.              The `decisive' distance is the second-largest effective distance              for the network, as detailed in the text. }
     \label{fig:all_sky_bns_fm}
\end{figure}

\Cref{fig:all_sky_bns_fm} shows the distribution of injections and their recovery as a function of the injected decisive distance and chirp mass.  We use these two parameters as they best 
encode the expected sensitivity of the search.  The overall amplitude of the \ac{GW} signal
scales with the chirp mass, $\mathcal{M} = m_{1}^{3/5} 
m_{2}^{3/5} (m_{1} + m_{2})^{-1/5}$, of the system.  The sensitivity of the search is typically
limited by the \ac{SNR} of the second most sensitive detector, and this is encoded in the
decisive distance.  The effective distance of a source is the distance at which an optimally-oriented 
and located signal would have the produced the same \ac{SNR} as the given simulation 
\cite{2008CQGra..25j5002B}.  The decisive distance is the second largest effective distance 
for the detectors in the network.  Given a requirement of power in at least two detectors, the
ability to detect an event will depend upon its decisive distance.

At lower masses, the majority of simulations injected below 30\,Mpc ($\sim 13$\,Mpc angle-averaged range) are successfully recovered, consistent with the network sensitive distance during \ac{S5} (\cref{fig:all_sky_range})\footnote{The angle-averaged range shown in \protect\cref{fig:all_sky_range} is calculated for a \protect\ac{BNS} with mass $m_1 = m_2 = 1.4\,\msun$, for which $\mathcal{M}_c \simeq 1.2$.}, with recovery improving as mass increases.
With the background highly cleaned by the myriad cuts and vetoes, resulting in a low-significance loudest event, very few simulations are marginally recovered, with the transition rapidly made to completely missed signals at higher distances.

\subsection{Comparison with the coincidence-based pipeline}

The recovery of injections as a function of distance allows the calculation of the volume 
sensitivity of the search.  This can then be compared with the coincidence search.
In order to perform the comparions, the same data segments were analysed using the 
coincidence-based \texttt{ihope} pipeline~\cite{Babak:2012zx}.
The same \acl{DQ} veto method was used, whereby those events vetoed in a single detector are removed from the search.
Finally, the search was performed using identical template parameters and simulation parameters, allowing a direct comparison of search efficiency with the new coherent pipeline.

\subsubsection{Computational cost}

The coherent and coincident algorithms are both limited by the speed of the \ac{FFT} --- the 
computational core of the matched-filter.
The coherent search has been implemented to ensure that the computation of the coherent SNR time
series, even when considering a large number of sky points, is dominated by the \ac{FFT}
computations needed to obtain the single detector SNR time series. Therefore, in Gaussian noise,
where coherent SNR is the optimal detection statistic, the computational cost of our coherent
search and a coincidence search would be roughly equivalent. However, to counteract
non-Gaussianities in the data we compute a set of \chisq statistics, which themselves include \ac{FFT}
computations. As the number of sky-points increases, the number of times the \chisq statistics
must be computed also increases, and for large sky grids this can be the dominant computational
cost. The coherent search has been implemented to ensure that the 
CPU-intensive calculation of the \chisq statistics is minimized by first applying single detector
thresholds and by applying the cheaper signal consistency tests---null SNR, auto $\chi^2$ and bank
$\chi^2$---before computing the expensive frequency-bin $\chi^2$.
Nevertheless, for the 2-site coherent search presented here we found that the algorithm
was roughly a factor of 2 more expensive than its coincidence-based predecessor due to additional
frequency-bin \chisq calculations. Recent work has demonstrated
that a non-\ac{FFT} based implementation of the frequency-bin $\chi^2$ can greatly reduce the
computational cost of that operation in the coincidence search \cite{Canton:2014ena}. We plan to
investigate whether a similar implementation can provide a similar improvement when applied
to the coherent search.

The implementation of background estimation via time shifts used in this analysis is
computationally costly.  Since we permute the order of the data segments, it is necessary
to re-compute the single detector \ac{SNR} time series for each time shift.  Thus, each time-slide 
is computationally equivalent to the zero-lag foreground, resulting in a further factor of ten increase
in computational cost for this search.  Performing time shifts in the coincidence analysis has
only a small impact in the total computational cost \cite{Babak:2012zx}.  So, with ten
time shifts, the coherent search is a factor of twenty more costly than the coincidence search. 
Of course, ten background trials is nowhere near sufficient to estimate event significance to 
detection level, where we might require a false rate of one per hundred or thousand years 
\cite{Colaboration:2011np}.  It will be a challenge to achieve this in a coherent analysis.
Recent developments \cite{Williamson:2014wma} have led to the implementation of shorter
time shifts which can be performed without the need to recalculate the single detector \ac{SNR}.
This allows for $\sim 30$ unique background trials at little additional cost.  Using this method, a
search including $1,000$ trials, using the current implementation,
would result in around a $\times 60$ computational cost in moving from coincidence-based to coherent.

\subsubsection{Signal recovery}

\begin{figure}
    \centering
    \includegraphics[width=.48\textwidth]{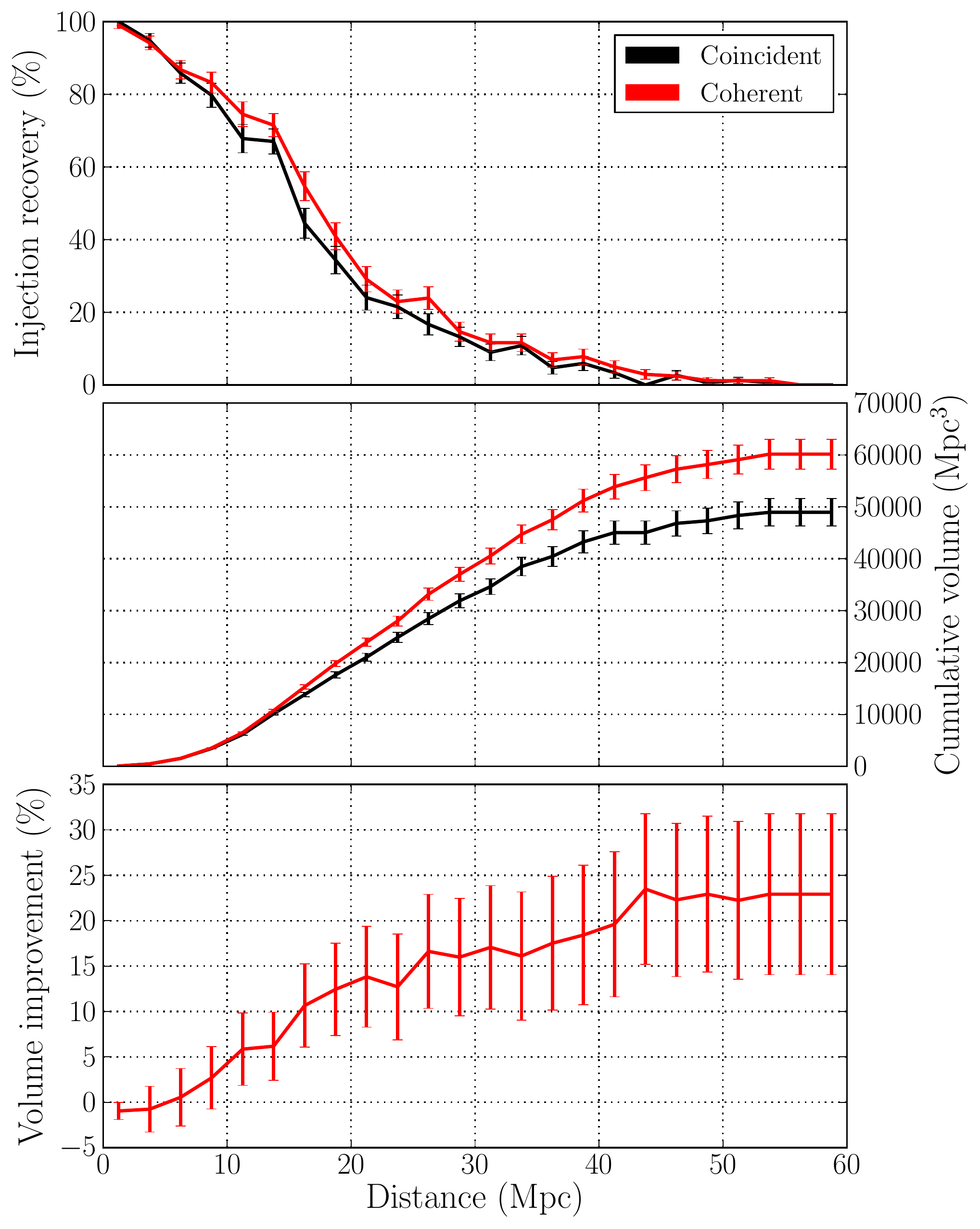}
    \caption[Comparison of search performance between the coincident and              coherent all-sky searches]            {Comparison of search performance between the coincident and              coherent all-sky searches for one month of S5 data.
             The top panel shows the injection recovery (efficiency) as a              function of distance, while the second panel shows the              the cumulative volume, comparing between the coincident (black)              and coherent (red) searches.
             The bottom panel shows the relative volume improvement of the              coherent search.
             The coherent search outperforms the coincident with nearly              25\% greater volume sensitivity.}
    \label{fig:all_sky_bns_efficiency}
\end{figure}

\Cref{fig:all_sky_bns_efficiency} compares the efficiency of simulation recovery between the two 
analyses. The top and centre panels compare the injection recovery and cumulative volume
respectively as functions of distance between the two searches; the bottom panel shows the
relative volume improvement of the coherent search. At low distances, the coherent
search recovers slightly fewer simulated signals than the coincident analysis, but the
differences are not statistically significant.  However, the near 25\% improvement in sensitive
volume over the full simulation campaign highlights the advantage of this coherent pipeline 
compared to the previously published algorithm.

We can compare the increased sensitivity of the coherent search with expectations.  
The coherent search employs a lower single detector \ac{SNR} threshold than the coincident 
search.  Specifically, the \ac{S5} search used a threshold of 5.5 in each detector, while we have 
required an \ac{SNR} above 5 in one detector and 4.5 in a second.  Furthermore, in the 
coherent analysis, the \ac{SNR} of the third detector will contribute to the coherent \ac{SNR}, regardless of its amplitude. In the coincident search, the \ac{SNR} of the third detector will only
 contribute if it is above threshold.  For the \ac{S5} data, where H2 was roughly half as sensitive as 
 H1 and L1, this means that a large fraction of signals will be below threshold in H2.  
To estimate the impact of these different thresholds, we generate a large number of simulated 
signals, uniformly distributed in volume, with uniform binary orientation.  For each, we calculate the 
expected \ac{SNR} in each detector (ignoring noise 
contributions) and count the number which would be ``detected'' by the coincident and coherent 
searches.  For the coherent search, we require the \ac{SNR} to be greater than 5 in one detector, 
and greater than 4.5 in a second, with the combined \ac{SNR} of 9.3 or more. In the coincident 
search, the \ac{SNR} in each detector must be above 5.5 before it contributes in the combined 
\ac{SNR}, which must be greater than 9.3.  We do not account for the discreteness of the template
bank, which will lead to a loss of \ac{SNR} due to a mis-match between the signal and template
waveforms.  Similarly, we neglect the loss in \ac{SNR} due to the discreteness of the sky grid
used in the coherent search.

The majority of events that pass the coherent search threshold have an \ac{SNR} below
5.5 in H2: 25\% are observed above threshold in H2, compared to over 90\% 
in both H1 and L1.  This means that only 25\% of the coherent sources are recovered as 
three detector coincidences. The remainder are observed in only two detectors (typically H1 
and L1, although there is a very small fraction that are
seen only in H1 and H2).  For these events, the \ac{SNR} from the third detector is not included in the coincident network \ac{SNR}, and consequently many of them will not be found above threshold.
Of the simulated events observable by the coherent search, only 80\% are observed by the coincident
search.  This is in excellent agreement with the results obtained by a full analysis on real data, where
we find the coherent analysis to be 25\% more sensitive.  
 \subsection{Future prospects}
\label{sec:future}

\begin{table*}
\caption{The relative performance of the coherent and coincident searches for various future
detector networks operating at their design sensitivity.  For each network, we consider a
detection threshold of 10 and 12.  For the coherent search, we require two detectors to
observe the signal above the SNR threshold of 5, but include the \ac{SNR} from \textit{all}
detectors.  For the coincident search, we only include the \ac{SNR} contribution from 
detectors where the signal would be above threshold.  The percentages give the sensitivity
relative to a search which imposes only a threshold on the network \ac{SNR}}
\label{tab:future_networks}
\centering
\begin{tabular}{|c|c|c||c||c|c|}
\hline
Network & Network & Sensitive Volume & Coherent Search
& \multicolumn{2}{c|}{Coincident Search}  \\
 & Threshold & ($10^{6} \mathrm{Mpc}^{3}$) & SNR 5 & SNR 4 & SNR 5.5 \\
\hline
\hline
\multirow{2}{*}{HLV} 
& 12 & 40 & $>99\%$ & $97\%$ & $90\%$ \\
& 10 & 65 & $>99\%$ & $94\%$ & $82\%$ \\
\hline
\multirow{2}{*}{HILV} 
& 12 & 55 & $>99\%$ & $94\%$ & $79\%$ \\
& 10 & 95 & $97\%$ & $87\%$ & $67\%$ \\
\hline
\multirow{2}{*}{HKLV} 
& 12 & 60 & $>99\%$ & $94\%$ & $80\%$ \\
& 10 & 103 & $96\%$ & $87\%$ & $67\%$ \\
\hline
\multirow{2}{*}{HIKLV} 
& 12 & 80 & $>99\%$ & $90\%$ & $70\%$ \\
& 10 & 135 & $93\%$ & $80\%$ & $57\%$ \\
\hline
\end{tabular}
\end{table*}

We have demonstrated the benefits of performing a fully coherent search and shown that
it leads to a 25\% increase in sensitivity for the H1-H2-L1 network that operated during the
initial detector era.  However, since the H2 detector does not form part of advanced LIGO,
this network will not operate in the future.  Indeed, it seems unlikely that there will be two
co-located detectors until the Einstein Telescope is operational \cite{Punturo:2010zz}.  
Prior to that, the coherent analysis presented here will need to be extended to a three (or more) site analysis before it will be useful.
Before undertaking this effort, it is worthwhile to investigate the likely benefits.

A planned evolution for the advanced LIGO and Virgo detectors is laid out at \cite{Aasi:2013wya}.  
The first
science run, expected in late 2015, will involve only the advanced LIGO detectors.  Following
that, the advanced Virgo detector will join the network with a sensitivity around a half of the 
advanced LIGO detectors during early runs, rising to two thirds when the detectors are operating
at design sensitivity at the end of the decade.  A third advanced LIGO detector, located in India 
\cite{Iyer:2011wb}, is expected to be operational around 2022 with a similar sensitivity to the other 
LIGO detectors.  The Japanese KAGRA \cite{PhysRevD.88.043007}
detector is being built and is expected to operate with a range similar
to that of the advanced LIGO detectors.  Therefore, the advanced detector network is expected
to evolve from a two detector network in the first run to a five detector network by early next decade.
Since the coherent analysis is not expected to benefit a two-detector search, we consider only 
networks of three or more detectors.  Concretely, we evaluate the benefit of the coherent analysis 
for the LIGO-Virgo (HLV), LIGO-KAGRA-Virgo (HKLV), LIGO (with India)-Virgo (HILV) and 
LIGO (with India)-KAGRA-Virgo (HIKLV) network.  

\Cref{tab:future_networks} provides the relative sensitivities of the coincident and coherent 
searches.  As before, we generate a large number of simulated signals and identify those which
would be observed by the searches.  Again, we ignore any loss in \ac{SNR} due to the
discreteness of templates in the mass space or points in the sky grid.
The sensitive volume available to each network is calculated assuming the only 
threshold is on the coherent network \ac{SNR}.  We evaluate the search sensitivity at a network
threshold of 12, corresponding to a very low false rate, as might be expected for first 
detections, and 10, which might be more realistic during the routine detection era.  For the 
coincident and coherent searches, we apply additional thresholds on the single detector
\acp{SNR} and we calculate the fraction of sources which would be
detected by the search.  In the coherent
search, we require an \ac{SNR} above 5 in at least two detectors for the event to be detected,
but the \ac{SNR} from \textit{all} detectors contributes to the network \ac{SNR}.  It is
clear from \cref{tab:future_networks} that this requirement has minimal impact on the
search sensitivity.  It is only for the five detector network that the single detector thresholds 
will reduce the detection rate by greater than 5\%, and these events would be recovered if we
were able to lower the single detector threshold to four.  

For the coincident search, we impose a single detector
threshold of 5.5 in each detector and only those detectors with a signal above threshold
contribute to the network \ac{SNR}.  This has a significant impact on the search sensitivity,
with 10\%/20\%/30\% of sources lost in a three/four/five detector search at SNR 12, increasing
to 20\%/30\%/40\% if the threshold can be lowered to 10.  As an alternative to implementing
a fully coherent analysis, one could simply lower the single detector threshold in the coincidence
search.  This would necessitate storing significantly more single detector triggers prior to 
performing
the coincidence step.  If the single detector threshold can be lowered to 4, then the majority of
sources are recovered.  Only with the five detector network, with observations being made 
at \ac{SNR} 10, do we lose 20\% of possible sources.  We note that requiring a larger SNR
threshold, say 5 or 5.5,
in two detectors may help to reduce noise background and will have minimal effect on signals.

There are two additional effects that must be taken into account when doing a careful 
comparison of searches: the computational cost of the searches, and their noise background.
The sensitivity of the coincidence search will be reduced due to the fact that the noise 
contributions from all detectors are incorporated into the network \ac{SNR}.  Thus, 
in Gaussian noise, the coherent search background will be $\chi^{2}$ distributed with
four degrees of freedom, while for the coincidence search it will be $\chi^{2}$ distributed with $2D$
degrees of freedom.\footnote{Since the searches impose single detector thresholds, the noise distribution will not be
exactly $\chi^{2}$ distributed, even in Gaussian noise.  Nonetheless, the background from the 
coincidence search will be elevated, relative to the coherent search.}
The impact of this is investigated in detail in \cite{DalCanton:thesis}, where noise background
for the coincidence search is shown to be several orders of magnitude higher 
that the coherent search, at a fixed \ac{SNR}.  Thus, a comparison of the searches
at \textit{fixed false alarm rate} requires a higher threshold on the coincidence search, 
further reducing the sensitivity. 
However, as discussed above, the computational cost of the coherent analysis will be higher 
than the coincidence search.  The computational cost of a search can be reduced by laying 
templates more sparsely in the mass-spin parameter space.  This will lead to a loss in sensitivity
as a signal is likely to have a poorer match with the closest template waveform.  
When computing resources are limited, comparison of the searches at 
\textit{fixed computational cost} would favour the coincidence
search.  However, without a full implementation of the coherent analysis, it is not possible to
perform the comparison.  We have argued that the current implementation of the coherent analysis,
with 1,000 background trials, is around 60 times that of the coincidence search.  Thus, strategies
to reduce the computational cost of the coherent analysis are required before we can make such a
comparison.

As an alternative to implementing the full coherent analysis, we could instead 
calculate the coherent \ac{SNR} for coincident events observed 
in three or more detectors.  The time delays between the detectors give a unique sky location (up to
a reflection symmetry in the three detector case), and this determines the detector
responses, $F^{X}_{+, \times}$.  Then, given the complex \ac{SNR}, $z^{X}$, from each detector,
we can calculate the coherent \ac{SNR} from \cref{eq:projection}, and also the null \ac{SNR} 
from \cref{eq:null_snr}.  This will reduce the noise background of the coincident search and provide
the null \ac{SNR} as an additional signal consistency test, and should 
yield many of the benefits of the coherent search, but with significantly lower overhead.  We note,
however, that the maxima of the single detector \acp{SNR} need not correspond to the maximum
coherent \ac{SNR}.  Additionally, for four or more 
detectors, the measured time delays may not be consistent with a physical sky location.  Thus,
it may be necessary to keep a short stretch of the \ac{SNR} time series around each trigger to
reconstruct the coherent \ac{SNR} \cite{2011CQGra..28m4009B}
and, at this stage, it may be easier to simply implement the 
coherent search.  Nonetheless, it is certainly worth investigating this approach, as it could
give a significant boost to the sensitivity of coincidence searches at minimal additional computational
cost.

\section{Discussion}
\label{sec:discussion}

We have demonstrated the first implementation of a fully-coherent all-sky search for 
\aclp{GW} from the inspiral of two compact objects.
This search extends the previously published targeted search for \ac{GW} signals associated 
with short \acp{GRB} \cite{Harry:2010fr, Williamson:2014wma} to the untargeted all-sky, all-time analysis. This fully-coherent, two-site search was seen to improve sensitive volume by nearly 
$25\%$ over a coincidence-based search of the same data.

We have argued that a similar improvement is to be expected for three detector networks in 
the advanced detector era.  The benefits for four and five detector networks are expected to be
even greater.  Nonetheless, the computational cost per template is significantly higher for 
the coherent analysis than the coincidence search.  Additionally, estimating the noise background 
through a time-shifted analysis of the data further increases the cost of the coherent analysis. It
will be difficult to obtain a background to ``detection level'' of one per hundred or thousand years
using this method, and alternatives \cite{Cannon:2012zt, Cannon:2015gha} may be needed.  
We have also
discussed methods by which the sensitivity of the coincidence search could be enhanced,
most notably by lowering the single detector thresholds (particularly on the least sensitive 
detectors), implementing a coherent follow up to all coincident events and incorporating
the null \ac{SNR}.  These possibilities deserve detailed investigation, in order to determine
the best way to implement a coherent analysis.  In addition, the search presented here made
use of much of the \texttt{ihope} infrastructure \cite{Babak:2012zx} used for the analysis of
initial LIGO and Virgo data.  In the meantime, there has been significant effort to modernise
and optimize the coincident analysis \cite{Canton:2014ena, Cannon:2011vi}.  Any coherent analysis of
the advanced detector data will have to build upon this new analysis infrastructure.

The first advanced detector runs will feature only the two LIGO detectors and, for this network, a coherent analysis is equivalent to the coincident analysis.  Nonetheless, the Virgo, KAGRA 
and LIGO India detectors will soon join the global network and, at this time, a coherent analysis
has the potential to significantly increase the rate at which gravitational wave signals from
binary mergers are observed.

 \section{Acknowledgements}
The authors would like to thank Patrick Sutton, Duncan Brown, B. Sathyaprakash, Valeriu Predoi, 
Sukanta Bose, Chad Hanna, Patrick Brady, Frank Ohme, Erin Macdonald, James Clark, Andrew
Williamson for discussions during the development
of the coherent analysis presented here.
In this work, D.~M.~M. was supported by the Science and Technology Facilities Council, UK, and by NSF award PHY-1104371.
SF would like to acknowledge
the support of the Royal Society and STFC grant ST/L000962/1.
IH acknowledges support from NSF awards PHY-0854812, PHY-0847611, PHY-1205835 and thanks the Max Planck Gesellschaft for support. 
\bibliographystyle{apsrev4-1}
\bibliography{../..//references/refs}

\end{document}